\documentclass[aps,twocolumn,showpacs]{revtex4}
\usepackage{amssymb}
\usepackage{latexsym}
\usepackage{amsmath}
\usepackage[mathcal]{eucal}

\usepackage{graphicx}
\usepackage{dcolumn}
\usepackage{amsfonts}
\usepackage{bm}
\usepackage{array}

\newcommand{\be}{\begin{equation}}
\newcommand{\ee}{\end{equation}}
\newcommand{\bea}{\begin{eqnarray}}
\newcommand{\eea}{\end{eqnarray}}

\renewcommand{\vec}[1]{{\bm #1}}

\bibpunct{[}{]}{,\!}{n}{,}{,} 

\begin{document}
\title{Collective modes in multi-band superconductors: Raman scattering in iron selenides}
\author{M. Khodas$^1$}
\author{A. V. Chubukov$^2$}
\author{G. Blumberg$^3$}

\affiliation{$^1$Department of Physics and Astronomy, University of Iowa, Iowa City, Iowa 52242, USA \\
$^2$Department of Physics, University of Wisconsin, Madison, Wisconsin 53706,
 USA\\
 $^3$Department of Physics and Astronomy
Rutgers, The State University of New Jersey, Piscataway, NJ 08854-8019 USA}

\pacs{74.20.Mn, 74.20.Rp, 78.70.Nx, 74.70.Xa}

\begin{abstract}
We study Raman scattering in the superconducting
state of  alkali-intercalated iron selenide
materials A$_x$Fe$_{2-y}$Se$_2$ (A=K,Rb,Cs)
in which Fermi surface has only electron pockets.
Theory predicts that both $s-$wave and $d-$wave pairing channels are attractive in this material,
and the gap can have either $s-$wave or $d-$wave symmetry, depending on the system parameters.
ARPES data favor $s-$wave superconductivity. We present the theory of  Raman scattering
 in A$_x$Fe$_{2-y}$Se$_2$  assuming that the ground state has s-wave symmetry but $d-$ wave is a close second.
 We argue that Raman profile  in $d-$wave $B_{2g}$ channel displays two collective modes. One is a particle-hole exciton, another is a Bardasis-Schrieffer-type mode
 associated with superconducting fluctuations in $d-$wave channel. At a finite damping, the two modes merge into one broad peak.
   We  present Raman data for A$_x$Fe$_{2-y}$Se$_2$ and compare them with theoretical Raman profile.
\end{abstract}
\pacs{74.25.nd,74.20.Rp,74.70.Xa}
\maketitle

\section{Introduction}
\label{Sec:Intro}

Superconductivity in Iron-based superconductors (FeSCs)  remains one of hottest topics in the  research on correlated electrons
~\cite{Mazin2010,Paglione2010,Johnston2010,Stewart2011,Basov2011,Wen2011,Hirschfeld2011,Chubukov2012,Yang2013,Shibauchi2014,Cvetkovic2013}.
The key issue, which is still under debates, is the symmetry of the superconducting order parameter  (OP) as it provides crucial input
  for microscopic description of FeSCs.
The phonon-mediated attraction is normally associated with a conventional $s$-wave pairing.
Alternative pairing mechanisms originating from the electron-electron interaction often give rise to non-s-wave pairing, like, e.g., $d-$wave pairing in the cuprates, but can also lead to an unconventional s-wave pairing in systems with multiple Fermi surfaces (FS) (Ref.\cite{maiti2013}). In the latter case, the gap is s-wave, but the OP changes sign across the Brillouin Zone (BZ).

Such an unconventional $s-$wave pairing state, often called $s^{+-}$, is believed to be realized in  weakly and moderately  hole and electron-doped Fe-pnictides, like Ba$_{1-x}$K$_{x}$Fe$_2$As$_2$, $x \lesssim 0.4$,  or Ba (Fe$_{1-x}$Co$_x$)$_2$As$_2$ \cite{Mazin2008,Kuroki2008,Graser2009,Hirschfeld2011,Chubukov2012} or in systems with isovalent substitution of one pnictide by the other, like BaFe$_2$(As$_{1-x}$P$_x$)$_2$ \cite{Yamashita2011,Shibauchi2014}.
The $s^{+-}$ superconductivity  is believed to originate from pair-hopping between electron and hole pockets, enhanced by spin-fluctiuations\cite{Mazin2008,Kuroki2008,Graser2009}.
 This pairing state is consistent with the number of experiments, including ARPES, neutron scattering, STM, NMR, optical conductivity and various thermodynamic measurements~~\cite{Paglione2010,Stewart2011,Chubukov2012,wen2012,Hirschfeld2011,Dusza2011,Valenzuela2013,Shibauchi2014,Evtushinsky2009a}.

Still, RPA-type~\cite{Graser2009,Maiti2011b} and Renormalization Group studies \cite{Chubukov2009,Platt2013,Maiti2013a} of weakly/moderately doped FeSCs show that there
are at least two attractive channels -- the attraction in  $s^{+-}$ channel is the strongest, but $d-$wave channel is also attractive, and the corresponding coupling is comparable to that in the $s^{+-}$ channel. Such close competition between the different pairing channels is ubiquitous in FeSCs and originates from the
  interplay between repulsive interaction between hole and electron pockets, which favors $s^{+-}$ superconductivity with the sign change of the gap between the two,
    and repulsion between, e.g., two electron pockets, which favors $d-$wave superconductivity with the sign change between the gaps on the two electron pockets~\cite{Graser2009,Kemper2010,Kuroki2009}  (by symmetry, the two electron pockets transform into each other under spatial rotation by $\pi/2$ around $z-$axis, and the $d-$wave gap changes sign under such a rotation).
    
   There is no theory restriction which would prevent $d-$wave attraction to become the strongest in some doping range.   For weakly/moderately doped FeSCs experimental data  seem to rule out $d-$wave superconductivity. However, at stronger doping, and, in particular, in systems with with only hole pockets or only electron pockets, the symmetry of the pairing state  is at the moment a highly controversial issue.
     The change of the pairing state upon doping would be quite interesting already on its own, but the interest is further triggered by the fact that the change from $s$ to $d-$symmetry can generate a mixed $s+id$ state in the intermediate doping range~\cite{Lee2009,Platt2012}. Such a mixed state breaks time-reversal symmetry and is highly sought  superconducting state as it has reach phenomenology~\cite{Nandkishore2012}.

For systems with only hole pockets, like strongly hole-doped KFe$_2$As$_2$,  functional RG calculations~\cite{Thomale2011a} favored the $d-$wave state, with the
    largest gap on the outer hole pocket, while RPA-type calculations~\cite{Maiti2011b,Maiti2012a} found near-identical couplings in $d-$wave and $s-$wave channels.
In the latter case
     the largest gaps are on the two inner hole pockets (the two $\Gamma-$centered pockets in Fe-only Brillouin zone (1FeBZ)). On experimental side,
      some thermal conductivity measurements  were interpreted~\cite{Reid2012a,Tafti2013} as strong evidence for $d-$wave pairing, while other thermal conductivity measurements~\cite{Shibauchi_private} and ARPES data for the same material~\cite{Ota2014} were interpreted as equally strong evidence for $s-$wave. Thermodynamic
  data were also interpreted~\cite{Hardy2013} as evidence for either $d-$wave or $s-$wave.

   For systems with only electron pockets, like  A$_x$Fe$_{2-y}$Se$_2$ Fe-selenides,
    RPA calculations within 5-band Hubbard-type model~\cite{Maier2011a,Das2011,Das2011a,Wang2011c,Maiti2011b}
    (the one which neglects doubling of the unit cell due to non-equivalent positions of Se compared to Fe plane) and fRG calculations~\cite{Wang2011c}
     yielded $d-$wave superconductivity due to a
     repulsion between electron pockets, while calculations within a metallic model with a purely magnetic spin-spin interactions between first and second neighbors (a metallic version of the $J_1-J_2$ model) yielded~\cite{Fang2011} a conventional $s-$wave pairing as in this model the interaction between electron pockets turns out to be attractive.
     On experimental side, ARPES experiments, particularly recent measurements of the superconducting gap along a small electron pocket centered at $k_z = \pi$ and $k_x = k_y =0$ in the actual (2Fe) zone\cite{Xu2012}, were interpreted as strong evidence for $s-$wave gap symmetry because the measured gap was argued to have only weak angular dependence, far from $|\cos 2 \theta|$, expected for a $d-$wave state.  At the same time, neutron scattering measurements on A$_x$Fe$_{2-y}$Se$_2$ showed\cite{Friemel2012,Friemel2012a} spin resonance in the superconducting
      state, which most, but not all~\cite{Saito2011}, researchers interpret as evidence for the sign change of the gap.
      Recently, two of us considered~\cite{Khodas2012a} the pairing in A$_x$Fe$_{2-y}$Se$_2$ within the model  which includes the hybridization between the electron pockets due to hopping via Se,   and found another $s^{+-}$  state,  in which the gap  changes sign between the hybridized bonding- anti-bonding electron pockets. This ``other $s^{+-}$''  state was originally proposed in \cite{Mazin2011}.  This state is $s-$wave, yet it supports spin resonance~\cite{Pandey2013}, in agreement with both ARPES and neutron scattering measurements.
      For repulsive interaction between electron pockets, this ``other'' $s^{+-}$ state competes with a $d-$wave state, and the winner of the competition is determined by the ratio of the hybridization and the (energy equivalent of) the ellipticity of the electron pockets~\cite{Khodas2012a}. 
If this ratio is small, $d-$wave wins, if it is large, $s^{+-}$ wins. 
In between, the system develops a mixed $s+id$ superconductivity at low temperatures.
For parameters relevant to A$_x$Fe$_{2-y}$Se$_2$, the ratio of  hybridization and ellipticity is of order one, and the couplings in $s^{+-}$ and $d-$wave channels are attractive and comparable in strength. 
In this respect, the situation at strong electron doping is quite similar to the one in strongly hole-doped materials.

 The presence of two different attractive channels in FeSCs and the uncertainty, both at the experimental and the theoretical level, about the pairing symmetry in systems with only hole or only electron pockets clearly calls for  measurements which can probe both pairing channels and, in particular, detect features associated with the subleading pairing channel, i.e., the one which does not cause superconductivity but is nevertheless an attractive one. The problem of this kind was considered by Bardasis and Schrieffer  (BS) back in 1961 (Ref.\cite{Bardasis1961}).  They argued that the subleading attractive pairing interaction gives rise to a
  collective mode  at an energy below $2\Delta$, where $\Delta$ is the superconducting gap generated by the primary pairing interaction.
   The presence of a collective mode below $2\Delta$ is the direct consequence  of residual attraction in this subleading channel.
   BS considered the case when the largest interaction is in $s-$wave channel and the gap $\Delta$ is a constant along the FS, but the analysis can be equally applied to cases when the leading pairing interaction  is in a channel with non-zero angular momentum. The only difference is that in this situation $\Delta$ has nodes and the excitonic BS mode should have a
    non-zero rate of damping into particle-hole continuum. 

 It has been argued~\cite{Lee2009,Scalapino2009} that that BS-type mode can be detected by Raman scattering, by analyzing Raman response in the subleading attractive channel.  A detection of the resonance in this channel at a finite energy below $2\Delta$ would indicate that (i) this channel is secondary and does not cause superconductivity and (ii) this channel is nevertheless an attractive one. Furthermore, the position of the peak would indicate to what extend this second channel is
 a competitor -- if the mode is close to $2\Delta$, the attraction in the secondary channel is weak compared to that in the leading channel, while if the mode frequency is near zero, the second channel is a strong competitor and can become the leading pairing channel upon a modest change of system parameters.

 For FeSCs with both hole and electron pockets present, the analysis of BS mode in the Raman profile has been presented in Ref. \cite{Scalapino2009}.
 It was argued that $B_{2g}$ Raman intensity should have a strong peak at a frequency of a BS collective mode 
(here and below we use the 2FeBZ  notations in references to Raman geometry).  The observation of BS mode has been reported by Kretzschmar {\it et. al.} in~\cite{Kretzschmar2013}.

In this communication we analyze the form of Raman profile in  A$_x$Fe$_{2-y}$Se$_2$ Fe-selenides.
 Like we said, these systems have only electron pockets, as  evidenced from both first-principle calculations~\cite{Zhang2009,Yan2011,Dagotto2013}
  and  ARPES measurements~\cite{Zhang2011,Mou2011,Qian2011,Wang2011,Zhao2011,Xu2012}).
    We assume that the interaction between the two electron pockets is repulsive.
  The pairing state in the absence of the hybridization between the electron pockets is $d-$wave, but the
    hybridization brings in a possibility for $s^{+-}$ superconductivity in which the gap is $s-$wave, but it changes sign between the two hybridized electron pockets~\cite{Mazin2011,Khodas2012a,Khodas2012}.
 The fact that the $s-$wave state emerges due to hybridization makes the analysis of the  Raman intensity in A$_x$Fe$_{2-y}$Se$_2$ Fe-selenides more involved compared to earlier analysis~\cite{Boyd2009,Scalapino2009,Chubukov2009} of Raman scattering in systems  with both hole and electron pockets, for which hybridization effects play little role and can be safely neglected.

    We assume, as ARPES data indicate, that the superconducting state in A$_x$Fe$_{2-y}$Se$_2$ has $s^{+-}$ symmetry and analyze Raman profile in $B_{2g}$ $d-$wave geometry. We show that in idealized situation of weak impurity-induced damping the $B_{2g}$ Raman intensity has two distinct near-$delta-$functional peaks.  
One peak is the BS mode caused by an attraction in the $d-$wave channel, the other is a particle-hole exciton, which exists because the $d-$wave density-density interaction is attractive.  The BS mode and particle-hole exciton are coupled, but we show that the coupling is parametrically weak in a $s^{+-}$ superconductor (the contributions  from the two pockets with different gap signs almost cancel each other, and the net result is non-zero only due to a finite ellipticity of electron pockets).  As a result, the two distinct peaks survive at small damping. At larger damping, the intensity fills in the region between the peaks and Raman intensity 
  acquires a shoulder-like form.  If the gap was a conventional, sign-preserving $s-$wave, the form of Raman profile would be very different as in this case the 
   coupling between BS mode and particle-hole exciton is strong  and only one combined peak develops below $2 \Delta$. 

We show that in our $s^{+-}$ case one of the two in-gap modes softens at the boundary between $s-$ and $s+id$ states.
This mode becomes indistinguishable in this limit from the original BS mode because the BS mode and the exciton in the particle-hole channel necessary decouple
 at zero frequency.  We show that the form of BS mode implies that the system develops $s+id$ and not $s+d$ order, i.e., it time-reversal symmetry gets broken 
 in the mixed state. 

\begin{figure}[h]
\begin{center}
\includegraphics[width=1.0\columnwidth]{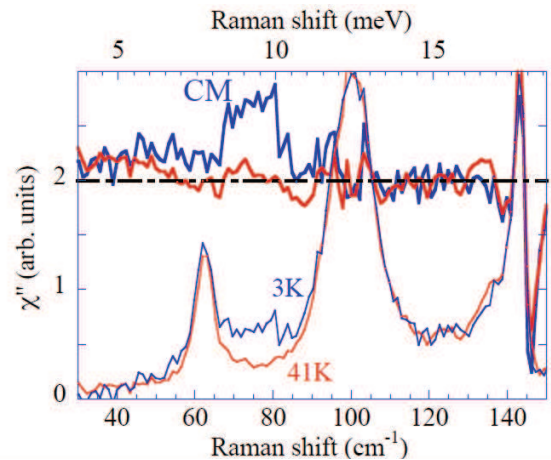}
\caption{
(color online) Low frequency Raman response from K$_{0.75}$Fe$_{1.75}$Se$_{2}$ superconductor in right-left scattering polarization channel in the normal state, at  $T=41$K (red), and in superconducting state at $T=3$K (blue),  from Ref.~\cite{Ignatov2012}. The upper panel shows the response with contributions from phonons subtracted,  as discussed in~\cite{Ignatov2012} .  The in-gap modes are found in the energy interval  8meV $\lesssim \omega \lesssim$ 11meV.}
\label{fig:data}
\end{center}
\end{figure}

We compare the structure of the theoretical $B_{2g}$  Raman intensity  below $T_c$ with the data for K$_{0.75}$Fe$_{1.75}$Se$_2$, reported in the Ref.~\cite{Ignatov2012}.
Fig.~\ref{fig:data} illustrates the Raman data for  in the  frequency interval where the enhancement of the Raman intensity below $T_c$ has been observed.
 We argue that the observed enhancement of the Raman intensity is consistent with the broadened double peak structure which we find theoretically.
 The in-gap modes are observed at $T=3$K in the interval 8meV $\lesssim \omega \lesssim$ 11meV.
 The ARPES measurements give $\Delta \approx 10$meV and $\Delta \approx 8$meV on a large Fermi surfaces according to Ref.~\cite{Xu2012} and Ref.~\cite{Borisenko2012} respectively.
 Ref.~\cite{Xu2012} gives $7$meV on a small symmetric $\kappa$ electron pocket.
These data places the energy of an in-gap modes observed in Ref.~\cite{Ignatov2012}  
well below  $2 \Delta$.
This indicates that the in-gap modes are strongly bound in K$_{0.75}$Fe$_{1.75}$Se$_2$, i.e., $d-$wave state is a strong competitor to $s-$wave state.

The paper is organized as follows.
In the Sec.~\ref{sec:Competing} we review the pairing scenarios in AFe$_2$Se$_2$ (A = Rb,K,Cs) compounds, and formulate the model for the description of our Raman data.
The calculation of the Raman intensity in $s^{+-}$ state is discussed in Sec.~\ref{sec:The_Raman}.
 We present our conclusions in Sec.~\ref{sec:Conclusions}

\section{Competition between  $s$ and $d$-wave orders in AFe$_2$Se$_2$ materials.}
\label{sec:Competing}
In this Section we review the theoretical  scenario which leads naturally to the competition between $s$-wave and $d$-wave pairing states.
  In Sec.~\ref{sec:Model} we describe the model in which the relative pairing strength in the two channels is controlled by the geometry of the electron pockets and the inter-pocket hybridization.
This model will also allow us to include density   fluctuations in the $d-$wave channel, which for brevity we will be calling the nematic fluctuations
We show that it is necessary to include these fluctuations to properly describe the Raman response.
Our goal is to describe the emergence of the strong Raman peak in $B_{2g}$ geometry  below $T_c$ for $s-$wave superconductivity.
We argue that $B_{2g}$ Raman peak is strong by two factors. First is
  proximity to the $d$-wave superconducting phase,  the second is the extra attraction provided by the nematic density fluctuations.

\subsection{The model}
\label{sec:Model}
We follow ~\cite{Khodas2012a} and consider the two-band model with generic short range interactions.
The model Hamiltonian  contains  the kinetic energy and the interactions.
The kinetic energy is quadratic in fermion operators and describes the excitations
 near the two Fermi pockets located at $(0,\pi)$ and $(\pi,0)$ in the 1FeBZ.
 We define $f^{\dag}_{1(2)\vec{k}}$  as the creation operator for electrons from the pocket at $(0,\pi)$ $[(\pi,0)]$,  and in each case count $\vec{k}$ as the momentum relative to the center of the corresponding pocket.
The quadratic part of the Hamiltonian $H=H_2 +H_{int}$ is
\begin{align}\label{H2}
H_2\! =\! \sum_{n=1,2}\!\sum_{\vec{k}}  \epsilon^{(n)}_{\vec{k}} f_{n\vec{k}}^{\dag} f_{n\vec{k}}
\!+\!
\sum_{\vec{k}} \!\lambda \!\left[ f_{1\vec{k}}^{\dag} f_{2\vec{k}}\!+\! f^{\dag}_{2\vec{k}} f_{1\vec{k}} \right] ,
\end{align}
 where the first term describes fermionic dispersion in 1FeBZ,
 and the second term  describes inter-pocket scattering with momentum transfer $\vec{Q} = (\pi,\pi)$.
 This second term hybridizes the two pockets. 
 It is allowed because the physical BZ is 2FeBZ due to two non-equvalent position of Se atoms staggered out of the Fe planes in a checkerboard fashion, \cite{Mazin2011,Khodas2012}.

For simplicity we neglect the out-of-plane dispersion, i.e., consider effective 2D problem.
Although such an approximation has to be applied with caution to describe finite momentum probes such as inelastic neutron scattering~\cite{Pandey2013}, we
 can safely use the 2D approximation to describe the zero momentum Raman response.

The simplest model dispersion yielding two elliptical FSs is
\begin{align}\label{Ec}
\epsilon^{(1,2)}_{\vec{k}} = \frac{ k_x^2  }{ 2 m_{x,y} } + \frac{ k_y^2  }{ 2 m_{y,x} }.
\end{align}
We set  $m_x < m_y$, in which case
  the Fermi pocket centered at $(0,\pi)$ has its major semi-axis along the $k_y$ axis.

The quartic interaction Hamiltonian is the sum of four terms allowed by symmetry:
\begin{align}\label{Hint}
H_1 & = \frac{u_1}{2}
\int d \vec{x}
\left(  f_{1\sigma}^{\dag} f_{2\sigma'}^{\dag} f_{2\sigma'} f_{1\sigma}
+ f_{2\sigma}^{\dag} f_{1\sigma'}^{\dag} f_{1\sigma'} f_{2\sigma}
\right)
\notag \\
H_2 & = \frac{u_2}{2}
\int d \vec{x}
\left(
  f_{1\sigma}^{\dag} f_{2\sigma'}^{\dag} f_{1\sigma'} f_{2\sigma}  +
   f_{2\sigma}^{\dag} f_{1\sigma'}^{\dag} f_{2\sigma'} f_{1\sigma}
   \right)
\notag \\
H_3 & =
\frac{u_3}{ 2 }
\int d \vec{x}
\left( f_{1\sigma}^{\dag} f_{1\sigma'}^{\dag} f_{2\sigma'}f_{2\sigma}   + f_{2\sigma}^{\dag} f_{2\sigma'}^{\dag} f_{1\sigma'} f_{1\sigma} \right)
\notag \\
H_4 & =
\frac{u_4}{2}
\int d \vec{x}
\left( f_{1\sigma}^{\dag} f_{1\sigma'}^{\dag} f_{1\sigma'} f_{1\sigma}
+  f_{2\sigma}^{\dag} f_{2\sigma'}^{\dag} f_{2\sigma'} f_{2\sigma}  \right)\, .
\end{align}
In Eq.~\eqref{Hint} $H_1$ and $H_2$ are inter-band density-density and exchange interactions,
 $H_4$ is the intra-band density-density interaction, and
 $H_3$ describes the umklapp pair-hopping processes.
The interactions with excess momentum $\vec{Q}$ do not play a role in the present analysis and we omit them.
For the underlying orbital model with local Hund and Hubbard interactions, $u_1 + u_2 = u_4 - u_3$ (Ref. \cite{Khodas2012a})
For simplicity, we assume that this condition holds. If it does not, the values of the couplings $u_d$ and $u_{\rho}$ in our consideration below will change, but the overall form of the Raman response will remain the same.

Two of us demonstrated in~\cite{Khodas2012a} that the superconducting OP in the model  specified by Eqs.~\eqref{H2} and \eqref{Hint} has $s$-, $d$- or $s+id$-symmetry depending on the ratio $\kappa = \lambda / \delta \epsilon $ of  the hybridization amplitude $\lambda$ and the  energy scale $\delta \epsilon$, related to ellipticity.
The latter is determined in the model by a typical energy separation, $\delta \epsilon_{\vec{k}}=\epsilon^{(1)}_{\vec{k}} - \epsilon^{(2)}_{\vec{k}}$ between the unhybridized pockets.
In explicit form, the parameter $\kappa$ is
\begin{align}\label{kappa}
\kappa = \frac{\lambda}{E_F}
\frac{ \mu_- }{ \mu_+ }\, ,\quad \mu^{-1}_{\pm} = \frac{1}{2}\left(m_x^{-1} \pm m_y^{-1}\right)\, .
\end{align}
The parameter $\kappa$ can be equally viewed as the ratio of the dimensionless hybridization $\lambda/E_F$ to the combination $\mu_+/\mu_-$, which characterizes the degree of pocket ellipticity.

\begin{figure}[h]
\begin{center}
\includegraphics[width=0.8\columnwidth]{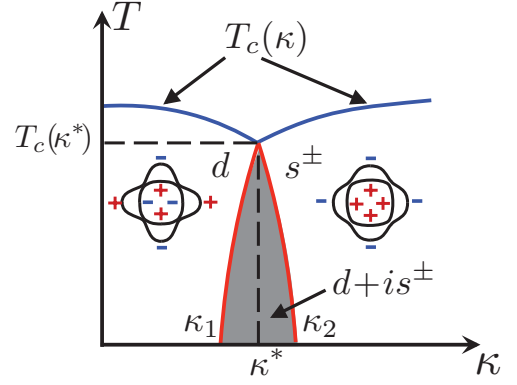}
\caption{ (color online)
Phase diagram of AFe$_2$Se$_2$ in $(\kappa,T)$ plane.
At high temperatures the system is in the normal state.
For $\kappa < \kappa^*$ ($\kappa > \kappa^*$)  the transition occurs to the $d$-wave $s^{+-}$ state.
The two normal to superconductor transitions merge at the tetra-critical point, $(\kappa^*,T^*)$.
For $T< T^*$ the system is in the $d$-wave state at $0< \kappa< \kappa_1^*(T)$,
in $s^{+-}$ state  for $\kappa > \kappa_2^*(T)$ and the intermediate $s+id$ phase with broken time reversal symmetry is obtained at $\kappa_1^*(T) < \kappa < \kappa_2^*(T)$.
The inset shows the energy of BS mode in the $s^{+-}$ state in units of the $s^{+-}$ OP, and the dashed line shows the minimal energy of quasi-particle excitations.
The BS mode softens closer to the transition to the $s+id$ state.
 }
\label{pd_BS}
\end{center}
\end{figure}

The phase diagram of the system is shown in Fig.~\ref{pd_BS}.
The OP just below $T_c(\kappa)$ has $s$($d$)-wave symmetry for $\kappa >(<) \kappa^*$.
Near $\kappa = \kappa^*$, there exists an interval $\kappa_1(T) < \kappa < \kappa_2(T)$ where the OP symmetry is $s+id$.
This interval extends from a point  at $T=T_c (\kappa^*)$,
 to a finite range $\kappa_1  < \kappa < \kappa_2$ at $T=0$.
 For our model with $u_1 + u_2 = u_4 - u_3$,  $\kappa^* = 1/ \sqrt{3}$.

The quadratic part of the Hamiltonian \eqref{H2} can be diagonalized by transforming to new fermionic operators  $a_{\vec{k}}$ and $b_{\vec{k}}$ satisfying
\begin{align}\label{ab}
a_{\vec{k}} &=  f_{1\vec{k}} \cos \theta_{\vec{k}} +  f_{2\vec{k}}\sin \theta_{\vec{k}}\, ,
\notag \\
b_{\vec{k}} & = - f_{1\vec{k}} \sin \theta_{\vec{k}} +  f_{2\vec{k}} \cos \theta_{\vec{k}}\, ,
\end{align}
where the angle of rotation in the orbital space is defined by
\begin{align}\label{angle}
\cos 2\theta_k & = \frac{\delta \epsilon_{\vec{k}} /2}{ \sqrt{ \lambda^2 + ( \delta \epsilon_{\vec{k}})^2/4 }}\, ,
\notag \\
\sin 2\theta_k &= \frac{\lambda}{ \sqrt{ \lambda^2 + ( \delta \epsilon_{\vec{k}})^2/4 }}\, .
\end{align}
The electron states created by operators $a_{\vec{k}}^{\dag}$ and $b_{\vec{k}}^{\dag}$ were termed anti-bonding and bonding states in Ref.~\cite{Hirschfeld2011} and we follow their notations.
In this work we focus on the domain $\kappa > \kappa_2(T)$ where the OP has an $s^{+-}$ symmetry.
The quasi-particle dispersion is determined by the eigenvalues of the
inverse Green function, which in the mean field approximation takes the form
 \begin{align}\label{matrix_G}
\hat{G}^{-1}_{\epsilon,\vec{k}} = \!\begin{bmatrix}
-i \epsilon + \xi_{\vec{k}}^a &  s_{\vec{k}} \Delta_s  & 0 &  c_{\vec{k}} \Delta_s \\
s_{\vec{k}} \Delta_s^*   & -i \epsilon - \xi_{\vec{k}}^a & c_{\vec{k}} \Delta_s^* & 0 \\
0 &  c_{\vec{k}} \Delta_s & -i \epsilon + \xi_{\vec{k}}^b &   - s_{\vec{k}} \Delta_s \\
c_{\vec{k}} \Delta_s^* & 0 &   - s_{\vec{k}} \Delta^*_s  & -i \epsilon - \xi_{\vec{k}}^b
\end{bmatrix},
\end{align}
where
$\xi_{\vec{k}}^{a,b}$ are the energies of bonding and anti-bonding states counted relative to the Fermi level, $E_F$,
\begin{align}\label{xi}
\xi_{\vec{k}}^{a,b} = & \frac{1}{2}\left(\epsilon^{(1)}_{\vec{k}} + \epsilon^{(2)}_{\vec{k}}\right)
- E_F
\notag \\
& \pm \frac{1}{2} \sqrt{\left(\epsilon^{(1)}_{\vec{k}} -\epsilon^{(2)}_{\vec{k}}\right)^2 + 4 \lambda^2}
\end{align}
In Eq. \eqref{matrix_G} we introduced shortened notations $c_{\vec{k}} = \cos 2 \theta_{\vec{k}}$,  $s_{\vec{k}} = \sin 2 \theta_{\vec{k}}$.

The matrix propagator in Eq.~\eqref{matrix_G} in general does not reduce to the block-diagonal form because of off-diagonal entries $c_{\vec{k}} \Delta_s$, which describe inter-pocket pairing of $a$ and $b$ fermions.
Such a pairing is contained in the term $\propto c_{\vec{k}} \Delta_s a_{\vec{k}}^{\dag} b_{\vec{k}}^{\dag}$ in the mean field Hamiltonian.
This term is allowed by symmetry, and  intra-band correlations $\propto c_{\vec{k}} \Delta_s a_{\vec{k}}^{\dag} b_{\vec{k}}^{\dag}$ are
 induced by proximity even when the superconductivity is driven by intra-band pairing~\cite{Rowe2013,Hinojosa2014}.
At the same time, the terms $\propto c_{\vec{k}} \Delta_s$ affect only states with momenta $\vec{k}$ such that $\xi_{\vec{k}}^a + \xi_{\vec{k}}^b \lesssim \Delta_s$. The momenta satisfying this condition fall in between of the two hybridized Fermi surfaces and are
 separated from the Fermi level by an energy of the order $\max\{\lambda,\delta \epsilon\}$. If $\Delta_s \ll \max\{\lambda,\delta \epsilon\}$,
  inter-pocket contributions are parametrically small compared to contributions from intra-pocket pairing terms in the Hamiltonian.
  To simplify presentation, we assume that the condition  $\Delta_s \ll \max\{\lambda,\delta \epsilon\}$ holds and neglect $c_{\vec{k}} \Delta_s$ terms in Eq.~\eqref{matrix_G}.  With this simplification, the mean-field Hamiltonian can be approximated by block-diagonal form
\begin{align}\label{matrix_G1}
\hat{G}^{-1}_{\epsilon,\vec{k}} \approx \!\begin{bmatrix}
-i \epsilon + \xi_{\vec{k}}^a &  s_{\vec{k}} \Delta_s  & 0 &  0 \\
s_{\vec{k}} \Delta_s^*   & -i \epsilon - \xi_{\vec{k}}^a & 0 & 0 \\
0 & 0 & -i \epsilon + \xi_{\vec{k}}^b &   - s_{\vec{k}} \Delta_s \\
0 & 0 &   - s_{\vec{k}} \Delta^*_s  & -i \epsilon - \xi_{\vec{k}}^b
\end{bmatrix},
\end{align}

It is convenient to introduce an extended Nambu notations,
\begin{align}\label{Nambu}
\chi^{\dag}_{\vec{k}} = & \left[\chi^{\dag}_{{\vec{k}}, a1}, \chi^{\dag}_{{\vec{k}},a2}, \chi^{\dag}_{{\vec{k}}, b1}, \chi^{\dag}_{{\vec{k}},b2} \right]
\notag \\
= & \left[a^{\dag}_{{\vec{k}} ,\uparrow}, a_{-\vec{k},\downarrow}, b^{\dag}_{{\vec{k}},\uparrow}, b_{-\vec{k},\downarrow}\right]\, .
\end{align}
The Pauli matrices  $\tau_{i}$, $i=1,2,3$ act on Nambu indices within each subband, and the other set of Pauli matrices $\varkappa_{i}$ with
 $i=1,2,3$ is operating in the space of the two subbands. 
 For block-diagonal structure of Eq.~\eqref{matrix_G1}, it's inverse in Nambu notations is
\begin{align}\label{matrix_G2}
\hat{G}_{\epsilon,\vec{k}}= G^{+}_{\epsilon,\vec{k}} + \varkappa_3 G^-_{\epsilon,\vec{k}}\, , \,\,\, G^{\pm}_{\epsilon,\vec{k}} = \frac{1}{2}\left(G^{(a)}_{\epsilon,\vec{k}} \pm G^{(b)}_{\epsilon,\vec{k}} \right)\, ,
\end{align}
where
\begin{align}\label{matrix_G3}
G^{(a,b)}_{\epsilon,\vec{k}} = \frac{i \epsilon + \tau_3 \xi^{a,b}_{\vec{k}}  \pm \tau_1 s_{\vec{k}} \Delta_s}{\epsilon^2 + \left(\xi^{a,b}_{\vec{k}}\right)^2 + s_{\vec{k}}^2 \Delta_s^2}\, .
\end{align}

\subsection{Raman susceptibility}
\label{sec:Raman}

The two photon Raman scattering cross-section  $I_R(\omega)$ is related to the imaginary part $\chi''(\omega)$ of the retarded Raman susceptibility, $\chi(\omega)$ by a standard relation
\begin{align}
I_R(\omega) = 2\left[ 1 + n_B(\omega)\right] \chi''(\omega)\,
\end{align}
with the Bose factor, $n_B(\omega) = \left(e^{\omega/T} - 1\right)^{-1}$.
The retarded Raman susceptibility,
\begin{align}\label{im_chi}
\chi(\omega) = - i \int_0^{\infty} d t \exp( i \omega t) \left\langle \hat{r}(t) \hat{r}(0) -\hat{r}(0) \hat{r}(t) \right\rangle \, ,
\end{align}
where $\hat{r}(t)$ is the Raman operator.  
In a general case~\cite{Klein1984,Devereaux2007,Chubukov1992}, the Raman operator contains two contributions -- the second order contribution associated
 with fermion current (the first derivative of the fermion dispersion over momentum) and the first-order contribution associated with the inverse effective mass (the second derivative of the dispersion over momentum). 
 The first contribution is important in the resonance regime, when the incoming fermionic frequency is adjusted to match a typical frequency of particle-hole excitations (Hubbard $U$ in case of Hubbard insulator) (Ref. \cite{Chubukov1992,Basko2008})
In the non-resonance regime, which we consider here, the second-order current contribution to the Raman vertex is not much different from the direct first-order contribution, and we can safely restrict with the inverse mass term. 
In this approximation, the Raman operator
\begin{align}\label{r}
\hat{r} = \sum_{\vec{k},i,j,(n)} \vec{e}^I_i M^{(n)}_{ij} \vec{e}^S_j
\end{align}
is determined by the polarization vectors of incoming and scattered photons, $\vec{e}^{I,S}$ and the effective mass tensor $M^{(n)}_{ij} = \partial^2 \epsilon^{(n)}(\vec{k})/ \partial k_i \partial k_j$ of an $n$th band, \cite{Klein1984,Devereaux2007}.
We focus on the $B_{2g}$ Raman configuration \cite{Boyd2009,Mazin2010a} relative to the (folded) 2FeBZ 
(which becomes $B_{1g}$ in the unfolded, 1FeBZ due to $45^o$ rotation between coordinate systems in the folded and unfolded zones).
The polarization vectors for $B_{2g}$ polarization are
$\vec{e}^{I,S} = (\hat{x} \pm \hat{y} )/ \sqrt{2}$ where the $\hat{x}$ and $\hat{y}$ are orthogonal unit vectors.
For the dispersion relation Eq.~\eqref{Ec}, we obtain from Eq.~\eqref{r}
\begin{align}\label{RB}
\hat{r}_{B_{2g}} = 2 {\mu}_-^{-1}
\sum_{\vec{k}}
\left( f^{\dag}_{1\vec{k}} f_{1\vec{k}} - f^{\dag}_{2\vec{k}} f_{2\vec{k}} \right)\, .
\end{align}
If the pockets were circular the $B_{2g}$ Raman response would vanish by symmetry.
At a non-zero ellipticity, this is no longer the case and $B_{2g}$ Raman intensity becomes finite.
 In the hybridized basis \eqref{ab}, the Raman vertex, \eqref{RB} takes the form
\begin{align}\label{RB1}
\hat{r}_{B_{2g}}\! \!=\! 2{\mu}_-^{-1}\!\sum_{\vec{k}} \!
\left[ \!c_{\vec{k}}\!
\left(\!\! a^{\dag}_{\vec{k}} a_{\vec{k}} \!- \!b^{\dag}_{\vec{k}} b_{\vec{k}} \right)
\!\!-\!\! s_{\vec{k}}\! \left( \!a^{\dag}_{\vec{k}} b_{\vec{k}} \!+\! b^{\dag}_{\vec{k}} a_{\vec{k}} \right)\!
\right]\!.
\end{align}
The condition $\Delta_s \ll \max\{\lambda,\delta \epsilon\}$ which allowed us to approximate Eq.~\eqref{matrix_G} by Eq.~\eqref{matrix_G1} also allows us to neglect inter-band contribution to the Raman vertex in Eq.~\eqref{RB1}, i.e., approximate $\hat{r}_{B_{2g}}$ by 
\begin{align}\label{RB2}
\hat{r}_{B_{2g}}\! \approx \!2 {\mu_-}^{-1}\sum_{\vec{k}} \!
c_{\vec{k}}\!
\left( a^{\dag}_{\vec{k}} a_{\vec{k}} \!- \!b^{\dag}_{\vec{k}} b_{\vec{k}} \right)\, .
\end{align}
The Raman vertex in Eq.~\eqref{RB2} describes the coupling of light to  $d$-wave 
 density  fluctuations.  
 The $d$-wave symmetry of the vertex Eq.~\eqref{RB2} is encoded in $c_{k_x,k_y} = -c_{k_y,-k_x}$.
Crucially, this Raman vertex Eq.~\eqref{RB2}  allows for the coupling to fluctuations of the $d-$wave superconducting OP. 
The coupling occurs via the triangular vertex which involves one normal and one anomalous Green function and one interaction line in $d-$wave particle-particle channel, see Fig.~\ref{fig:diagrams}.
  This  triangular vertex does not vanish by symmetry because both $\hat{r}_{B_{2g}}$ and $s^{+-}$ gap change sign between the hybridized bands.
 The coupling to $d-$wave particle-particle channel give rise to BS modes, as we discuss in the next section.

\section{The Raman intensity in $s^{+-}$ state}
\label{sec:The_Raman}

In this section we calculate the Raman intensity, Eq.~\eqref{im_chi} assuming that the superconducting state has $s^{+-}$ symmetry.

Equations \eqref{im_chi} and \eqref{RB2} show that the Raman intensity is determined by the correlation function of the $d$-wave density operator, $c_{\vec{k}}(a_{\vec{k}}^{\dag} a_{\vec{k}} - b_{\vec{k}}^{\dag} b_{\vec{k}})$, which in Nambu notations, Eq.~\eqref{Nambu}, reads
\begin{align}\label{d_rho}
c_{\vec{k}}\sum_{\sigma}(a_{\vec{k}\sigma}^{\dag} a_{\vec{k}\sigma} - b_{\vec{k}\sigma}^{\dag} b_{\vec{k}\sigma})
= c_{\vec{k}} \chi^{\dag}_{\vec{k}}\left[ \tau_3 \varkappa_3 \right] \chi_{\vec{k}}\, ,
\end{align}
where, as before, $\tau_{i}$, $i=1,2,3$ are Pauli matrices acting on Nambu indices within each subband, and $\varkappa_{i}$, $i=1,2,3$ operate in the space of the two subbands.

To leading (zero) order in the interaction, $\chi''(\omega)$ is proportional to the convolution of the two fermionic propagators with $d-$wave vertices. 
Interactions leads to two types of effects. First, $d-$wave particle-hole vertex gets dressed by $d-$wave density-density interaction. If this interaction is 
 attractive, one can expect an exciton-like resonance below $2\Delta$, where $\Delta$ is $s^{+-}$ gap.  Second, a triple vertex which we discuss at the end of  previous section converts $d-$wave particle-hole propagator into $d-$wave particle-particle propagator. The latter than gets dressed by the $d-$wave interaction on the particle-particle channel. If the latter is attractive, one can expect another resonance below $2\Delta$, which is a $d-$wave BS mode. 
 As a result, Raman profile below $2\Delta$ can have two peaks.  In the presence of impurity scattering, the two peaks gets broadened, and one should generally expect
  Raman intensity to get enhanced in a finite frequency range below $2\Delta$.  
  
 We show below that this is indeed what we obtain in the calculations. 
 Before we proceed, we note that, in general, there can be two resonance modes in the particle-particle channel, one is associated with the longitudinal fluctuations of the $d-$wave superconducting order parameter, another is associated with phase fluctuations.  
Let us assume for definiteness that the ground state $s^{+-}$ OP is real.
Then the two modes describe fluctuations of the real and the imaginary part of the $d$-wave OP.
This was realized already by BS.
In the present context the collective variables describing these two modes of OP oscillations are
\begin{subequations}\label{d_Cooper}
\begin{align}\label{Lambda}
c_{\vec{k}}\!\!
\left[
( a_{\vec{k}\downarrow} a_{\vec{k}\uparrow}\! + \!
a_{\vec{k}\uparrow}^{\dag} a^{\dag}_{\vec{k}\downarrow})
\!-\!
(a \!\rightarrow\! b) \!\right]
\!=\!
c_{\vec{k}} \chi^{\dag}_{\vec{k}} \left[\tau_1\varkappa_3 \right] \chi_{\vec{k}}
\end{align}
 \begin{align}\label{Gamma}
ic_{\vec{k}}\!\!
\left[\!
(a_{\vec{k}\downarrow} a_{\vec{k}\uparrow}  \! - \! a_{\vec{k}\uparrow}^{\dag} a^{\dag}_{\vec{k}\downarrow})
\!-\!
(a \!\rightarrow\! b) \!\right]
\!=\!
c_{\vec{k}} \chi^{\dag}_{\vec{k}} \left[\tau_2  \varkappa_3 \right] \chi_{\vec{k}}
\end{align}
\end{subequations}
Note similar structure of the operators \eqref{d_rho} and \eqref{d_Cooper}.
We define the matrix correlation function with entries
\begin{align}\label{chi_matrix}
\hat{\chi}_{\alpha\beta} = \left\langle \hat{\tau}_{\alpha} \hat{\tau}_{\beta} \right\rangle_{\omega}
\end{align}
defined as Matsubara Green functions of collective variables,
\begin{align}\label{tau_coll}
\hat{\vec{\tau}} = \sum_{\vec{k}} c_{\vec{k}} \chi^{\dag}_{\vec{k}} \left[\vec{\tau}  \varkappa_3 \right] \chi_{\vec{k}}
\end{align}
The Raman susceptibility, Eq.~\eqref{im_chi} is
\begin{align}\label{im_chi1}
\chi''(\omega) = \mathrm{Im}\left[\hat{\chi}_{33}(\omega)\right]\, .
\end{align}

To compute $\chi''(\omega)$ we 
 project the interaction Hamiltonian, Eq.~\eqref{Hint}, on the $s$- and $d$-wave Cooper channel and the $d$-wave density channel.
\begin{align}
H_{int} \approx V_{C}^{s} + V_{C}^{d} + V_{\rho}^d\, .
\end{align}
Keeping only the parts of the interaction Hamiltonian which contain intra-band processes, we obtain
\begin{subequations}\label{channels}
\begin{align}\label{V_C_s}
V_{C}^{s} =& - \frac{u_s}{2} \sum_{\vec{k},\vec{k}'}\sum_{\sigma,\sigma'}
s_{\vec{k}}\left[ a^{\dag}_{\vec{k},\sigma} a^{\dag}_{-\vec{k},\sigma} -    b^{\dag}_{\vec{k},\sigma} b^{\dag}_{-\vec{k},\sigma}
\right]
\notag \\
& \times c_{\vec{k}'}\left[ a_{\vec{k}',\sigma'} a_{-\vec{k}',\sigma'} -    b_{\vec{k}',\sigma'} b_{-\vec{k}',\sigma'}
\right]\, ,
\end{align}
\begin{align}\label{V_C_d}
V_{C}^{d} =& - \frac{u_d}{2} \sum_{\vec{k},\vec{k}'}\sum_{\sigma,\sigma'}
c_{\vec{k}}\left[ a^{\dag}_{\vec{k},\sigma} a^{\dag}_{-\vec{k},\sigma} -    b^{\dag}_{\vec{k},\sigma} b^{\dag}_{-\vec{k},\sigma}
\right]
\notag \\
& \times c_{\vec{k}'}\left[ a_{\vec{k}',\sigma'} a_{-\vec{k}',\sigma'} -    b_{\vec{k}',\sigma'} b_{-\vec{k}',\sigma'}
\right]\, ,
\end{align}
\begin{align}\label{V_rho_d}
V_{\rho}^{d} =&- \frac{u_{\rho}}{2} \sum_{\vec{k},\vec{k}'}\sum_{\sigma,\sigma'}
c_{\vec{k}}\left[ a^{\dag}_{\vec{k},\sigma} a_{\vec{k},\sigma}  -    b^{\dag}_{\vec{k},\sigma} b_{\vec{k},\sigma}
\right]
\notag \\
& \times c_{\vec{k}'}\left[ a^{\dag}_{\vec{k}',\sigma'} a_{\vec{k}',\sigma'} -    b^{\dag}_{\vec{k}',\sigma'} b_{\vec{k}',\sigma'}
\right]\, .
\end{align}
\end{subequations}
The interaction amplitudes are (see \cite{Khodas2012a} and Appendix ~\ref{app:Interaction})
\begin{align}\label{amplitudes}
u_d &= \frac{1}{2}(u_3 - u_4)\, , u_s = -\frac{1}{2}(u_1+u_2)
\notag \\
u_{\rho} &= u_1 - \frac{1}{2}(u_2+u_4)\, .
\end{align}
The interactions in the $d$-wave channel, Eqs.~\eqref{V_C_d} and \eqref{V_rho_d}, can be conveniently rewritten in terms of the collective variables introduced in Eq.~\eqref{tau_coll} as
\begin{align}
V_{C}^{d} = - \frac{u_d}{2} \hat{\tau}_+\hat{\tau}_-\, ,
V_{\rho}^{d} = - \frac{u_d}{2} \hat{\tau}_3\hat{\tau}_3\, ,
\label{aca}
\end{align}
where $\hat{\tau}_{\pm} = \hat{\tau}_1 \pm i \hat{\tau}_2$.

In our case the ``amplitude'' mode, Eq.~\eqref{Lambda} is coupled neither to the ``phase''  modes nor to density fluctuations in the $d$-wave channel, Eq.~\eqref{RB}, and therefore does not show up in the Raman response. 
We therefore can safely neglect the ``longitudinal'' mode of $d-$wave OP and truncate the matrix Eq.~\eqref{chi_matrix} to a two-by-two matrix with indices
$\alpha,\beta = 2,3$.
Projecting  the amplitude mode simplifies the $d$-wave Cooper channel to $V_{C}^{d}$ in (\ref{aca}) to  $V_{C}^{d}= - (u_d/2) \hat{\tau}_2\hat{\tau}_2$.

\begin{figure}[h]
\begin{center}
\includegraphics[width=0.99\columnwidth]{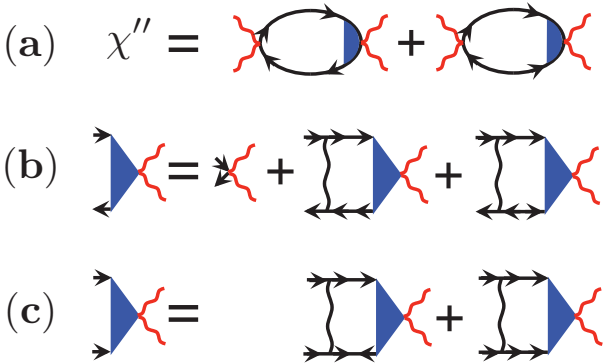}
\caption{ (color online)
(a) Diagramatic representation of the Raman susceptibility, $\chi''$.
(b) p-h and (c) Cooper Raman vertices within the ladder approximation. 
Vertex renormalizations (often called final state interaction in Raman literature) represent
multiple interactions in particle-hole channel and processes which convert particle-hole into particle-particle channel and include multiple scattering events in the particle-particle channel.  
 }
\label{fig:diagrams}
\end{center}
\end{figure}

The full Raman intensity is obtained by combining the processes with multiple interactions in particle-hole channel and processes which convert particle-hole into particle-particle channel and include multiple scattering events in the particle-particle channel.
We compute $\chi'' (\omega)$ by summing up series of ladder diagrams in the particle-particle and particle-hole channel. 
The corresponding diagrams are shown in Fig. \ref{fig:diagrams}.

In analytical form we have
\begin{align}\label{chi_full}
\hat{\chi}=
\hat{\Pi}
\left( \hat{I} + \hat{V} \hat{\Pi} \right)^{-1}\, ,
\end{align}
where $\hat{\Pi}$ is the non-interacting $\hat{\chi}$ matrix,
$\hat{I}$ is the two-by-two unit matrix, and the matrix
\begin{align}\label{V}
\hat{V} = \frac{1}{2}
\begin{bmatrix}
u_d & 0 \\
0 & u_{\rho}
\end{bmatrix}\, .
\end{align}
Equations \eqref{chi_full}, \eqref{V} give for the Raman susceptibility, Eq.~\eqref{im_chi1}
\begin{align}
\label{I_R}
\chi''(\omega) = \mathrm{Im} \left[ \frac{ -u_d \Pi_{32} \Pi_{23}/2  + \Pi_{33} (1 + u_d \Pi_{22}/2) }{ ( 1 + u_d\Pi_{22}/2 )( 1 + u_{\rho} \Pi_{33}/2 ) - u_d u_{\rho} \Pi_{23} \Pi_{32}/4 } \right]\, .
\end{align}
The matrix elements of the polarization operator are
\begin{align}\label{Pi}
\Pi_{ij}(\omega)\! =\! \!\int \!\!\frac{ d \epsilon}{ 2 \pi }\! \int \!\!\!\frac{d \vec{k}}{(2 \pi)^2}
\!\mathrm{Tr}\! \left[ \hat{G}(\epsilon\! +\! \omega,\vec{k}) \hat{\tau}_i  \hat{G}(\epsilon, \!\vec{k}  ) \hat{\tau}_j \right].
\end{align}
In Eq.~\eqref{Pi} the Green function is defined in Eq.~\eqref{matrix_G1} and the trace is taken over the extended Nambu indices.
We present the details of the calculation of the elements of Eq.~\eqref{Pi} 
in Appendix~\ref{app:Pi}, and here  quote the result:
\begin{subequations}\label{pol_op}
\begin{align}\label{pol22}
\Pi_{22} =
&- \frac{2}{ u }\left[  \frac{ \langle c^2 \rangle}{ \langle s^2 \rangle} \right]
-2 \frac{ \langle c^2\rangle}{ \langle s^2\rangle} \langle s^2 \log s^2 \rangle + 2\langle c^2 \log s^2 \rangle
\notag \\
& -
4 x \left\langle  \frac{ c^2 \arcsin \left(x/s\right)}{\sqrt{s^{2}-x^2 }} \right\rangle
\end{align}
\begin{align}\label{pol23}
\Pi_{23}= \Pi_{32}^*
=
- 4 i  \left\langle \mathcal{F}(\mu_+/\mu_-;\kappa,\phi) \frac{   \arcsin(x/s) }{ \sqrt{s^2 - x^2 }} \right\rangle
\end{align}
\begin{align}\label{pol33}
\Pi_{33}  =
 - 4 \left\langle
\frac{ c^2 s^2 \arcsin \left( x/s \right)}
{x \sqrt{s^2 - x^2}}
\right\rangle \, .
\end{align}
\end{subequations}
In Eqs.~\eqref{pol_op} we use the dimensionless variable $x = \omega/(2\Delta_s)$ and the angular brackets, $\langle \ldots \rangle$ indicate averaging over the directions of the vector $\vec{k}$ specified by the angle $\phi$ which vector $\vec{k}$ forms with the $k_x$-axis in the BZ.

The magnitude of the off-diagonal polarization operator, Eq.~\eqref{pol23} is determined by the dimensionless function $\mathcal{F}$ which depends on the interplay between superconducting gaps on the bonding and anti-bonding Fermi surfaces. 
 For a conventional sign-preserving superconducting OP $\mathcal{F} =1$. In our case, $\mathcal{F}$ is strongly reduced  
 To see this we note that $\Pi_{23}$ contains products of the normal and anomalous Green functions and is therefore an odd function of the  the $s^{+-}$ OP, $\Delta_s$.
 The contributions from the bonding and anti-bonding bands then have opposite signs and tend to cancel. The cancellation would be exact if the Fermi pockets were
  circular. In our case of elliptical pockets, the cancellation is not complete and in the limit of 
 weak ellipticity and hybridization, $\left\{ \lambda/E_F, \mu_+/\mu_-\right\} \ll 1$ but $\kappa = O(1)$,  we obtain (see Appendix~\ref{app:Pi23} for details)
\begin{align}\label{F_calig1}
{\cal F}\left(\frac{\mu_+}{\mu_-} \ll 1;\kappa,\phi \right) \approx   \frac{\mu_+}{\mu_-}
\left( 2c_{\vec{k}}^2 s_{\vec{k}}^2 - c_{\vec{k}}^4(1 + \kappa)    \right)\, .
\end{align}
The proportionality of $\mathcal{F}$ to $ {\mu_+}/{\mu_-}$ is the key result here. 
We see that the term, which mixes contributions from particle-hole and particle-particle channels, is parametrically small for $s^{+-}$ superconductivity and near-circular pockets.   
The angle-dependent term in  \label{F_calig1}  is not important as it yields $O(1)$ after angular integration. By this reason, in numerical calculations below we approximate ${\cal F}\left(\frac{\mu_+}{\mu_-} \ll 1;\kappa,\phi \right)$ by
\begin{align}\label{F_calig1a}
{\cal F}\left(\frac{\mu_+}{\mu_-} \ll 1;\kappa,\phi \right) \approx   \frac{\mu_+}{\mu_-}.
\end{align}

In the limit of strong ellipticity, ${\mu_+}/{\mu_-} = O(1)$ and $\mathcal{F}$ is a non-universal number of order one.

It is clear from Eq.~\eqref{I_R} that the Raman susceptibility is peaked at the frequencies where the denominator in Eq.~\eqref{I_R} vanishes. 
In the absence of the coupling between particle-hole and particle-particle channels, i.e., at $\Pi_{23} = \Pi_{32} =0$, the two poles in $\chi''$ at $\omega < 2\Delta$  would correspond to two distinct collective modes -- a  BS mode at  $1 + u_d \Pi_{22}/2 =0$ and a particle-hole  exciton at $1 + u_\rho \Pi_{33}/2 =0$. 
In both cases, to obtain the corresponding mode one needs an attractive interaction. 
In our case, both $u_d$ and $u_\rho$ are positive, i.e., both collective excitations are present and are Raman-active. 
The existence of Raman-active particle-hole excitons in Fe-pnictides is not new -- earlier an $s-$wave particle-hole exciton was argued to be present in $A_{1g}$ Raman channel in systems with both hole and electron pockets~\cite{Chubukov2009}.
   
At a non-zero $\Pi_{23}$, the two modes get coupled, but, as long as the coupling is small and the mode frequencies are at some finite distance from each other, the two-pole structure of $\chi''(\omega)$ at $\omega < 2 \Delta$ survives, although each collective excitation  becomes a mixture of an exciton and a BS mode.   
 
\begin{figure}[h]
\begin{center}
\includegraphics[width=0.99\columnwidth]{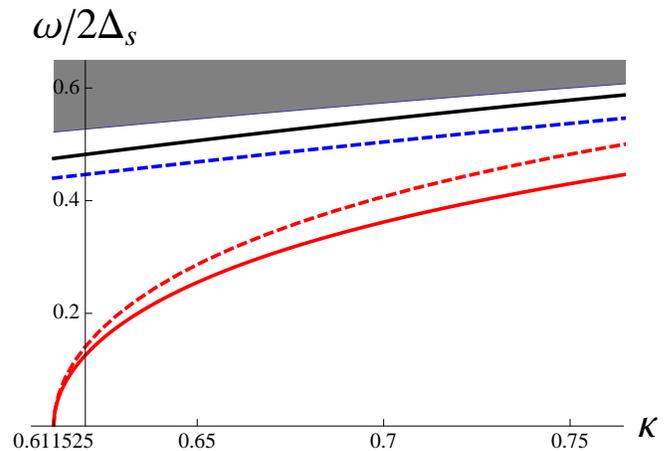}
\caption{ (color online)
B$_{2g}$ collective modes (solid lines) of an $s^{+-}$ superconductor at $T=0$ (the superconducting gap changes sign between the two hybridized Fermi pockets). 
 The energies of the collective excitations are plotted 
  as functions of the dimensionless parameter $\kappa$ introduced in Eq.~\eqref{kappa}.
 The energies are in units of $2 \Delta_s$, where 
  $\Delta_s$ is the magnitude of $s^{+-}$ superconducting OP.
The modes are shown in the interval  $\kappa>\kappa_2 \approx 0.611525$ (see Fig. \ref{pd_BS}). 
At the energies of the two modes  the denominator in Eq.~\eqref{I_R} vanishes,  giving rise to the delta-functional peak in the Raman intensity $\chi''$.
We used  Eqs.~\eqref{pol_op} for the polarization operators $\Pi_{ii}$ in Eq.~\eqref{I_R} and used the approximate form for $\Pi_{23}$, \eqref{F_calig1a}.
The parameters used in the calculation are $u_d = 0.4$,  $u_{\rho}=0.6$, $\mu_+/\mu_- = 0.15$.
The lower (red)  and upper(blue) dashed lines are obtained in the limit $\Pi_{23} = \Pi_{32} = 0$, and represent the BS mode and a particle-hole exciton, respectively.
 The coupling between the two channels mixes the BS mode and p-h exciton.
This coupling is, however, weakened for $s^{+-}$ gap superconductivity and scales with the degree of ellipticity of electron pockets. 
The shaded area is the quasi-particle continuum whose lower boundary defined by $2 \Delta_s (\kappa/\sqrt{\kappa^2 +1})$.
This boundary approaches $2 \Delta_s$ for large hybridization, $\kappa \gg 1$.}
\label{fig:modes}
\end{center}
\end{figure}

In Fig.~\ref{fig:modes} we show the behavior of the two modes as a function of $\kappa$ with and without the mixing term. The upper mode is predominantly an exciton, the lower one is a BS mode. The two modes repel each other, as it is expected as the ``coupling term'' in Eq.~\eqref{I_R} is repulsive.

The frequencies of the modes in the two channels as well as the energies of the actual, coupled excitations are shown in Fig.~\ref{fig:modes} as a function of the parameter $\kappa$.
These results are obtained by numerically performing angular integrations in the Eqs.~\eqref{pol22}, \eqref{pol23} and \eqref{pol33} and finding the roots of the equation $\det[\hat{\chi}(x)] = 0$.
The BS mode  softens when the parameter $\kappa$ decreases towards the critical value $\kappa=\kappa_2$ and the system undergoes the transition from $s^{+-}$ to $s+id$  superconductor. 
We emphasize that  the ``phase'' mode rather than the ``amplitude'' mode becomes critical.  The ``phase'' excitations are in the direction transverse to 
 the direction of the phase of the  $s^{+-}$  OP. 
Hence a condensation of the phase mode implies that the resulting state is $s +id$. This is consistent with the GL analysis in~\cite{Khodas2012a}.  
If, instead, longitudinal mode would soften, the resulting state would be $s+d$.  
We also note that the transition  from $s$ to $s+id$ at $\kappa=\kappa_2(T=0)$  breaks a discrete time reversal symmetry (an Ising-type transition) and therefore does not lead to the appearance of a Goldstone mode. 
As a result, the BS mode must bounce back to a finite value at $\kappa < \kappa_2$. 
Finally, we note that the softening of the BS mode is not affected by the particle-hole exciton. 
Combined mode softens  because the BS mode and the exciton decouple at $\omega =0$. Indeed, one can easily find from \eqref{pol23} that $\Pi_{23}(x=0) =0$. 
From physics perspective, the vanishing of the coupling is the consequence of the fact that the phase of a superconducting OP enters the quantum action only via spatial or temporal derivatives and hence the coupling between the phase mode and other modes must vanish at zero frequency. 
 
\begin{figure}[h]
\begin{center}
\includegraphics[width=1.0\columnwidth]{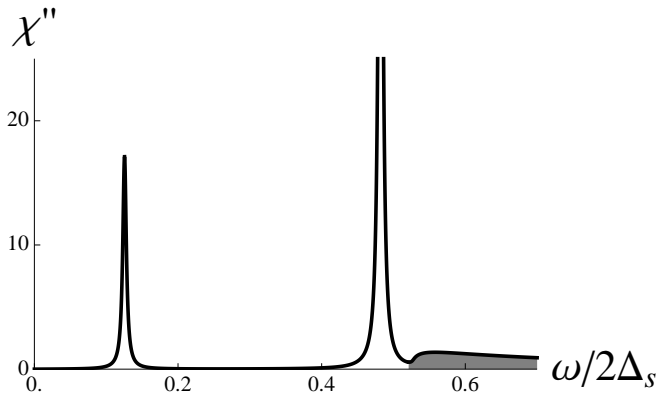}
\caption{(color online)
Calculated Raman susceptibility, $\chi''$ as a function of the dimensionless frequency, $\omega/ 2 \Delta_s$ for the fixed $\kappa=0.62 > \kappa_2$.
The two peaks represent in-gap modes corresponding to the two solid lines in Fig.~\ref{fig:modes}.
The parameters are the same as in Fig.~\ref{fig:modes}.
The continuum starts once the frequency enters the shaded area in Fig.~\ref{fig:modes}. 
The small imaginary part $0.003(2 \Delta_s)$ was added to the frequency for regularization.}
\label{fig:Raman_1}
\end{center}
\end{figure}

The Raman susceptibility calculated by substitution of Eq.~\eqref{pol_op} into Eq.~\eqref{I_R} is shown in Fig.~\ref{fig:Raman_1}.
In an idealized case of vanishingly small damping, the Raman intensity contains  two nearly delta-functional peaks, the lower one is predominantly a BS mode, the upper one is predominantly an exciton in the particle-hole channel.  
At higher degree of disorder, the peaks get broader and intensity in the region between the peaks  gets increased. 

Figure \ref{fig:exp} illustrates this build up of the Raman intensity for the specific choice of parameters.
The peaks at lower and higher energies represent the BS  and exciton modes respectively the same way as in Fig.~\ref{fig:Raman_1}.
Except in the immediate vicinity of the boundary between the $s^{+-}$ and $s+id$ phases the two modes have roughly similar binding energies of the order of $\Delta$, see Fig.~\ref{fig:modes}.
The mixing between the two channels tend to repel the two modes in frequency similar to a familiar level repulsion, which is also shown in Fig.~\ref{fig:modes}.
In the case of $s^{+-}$ OP however such a mixing is a weak effect.
And the peaks will in general stay close in energy.

In the presence of the disorder and inhomogeneous broadening the frequency interval between the two peaks is filled up below the superconducting transition.
This trend is in agreement with the experimental results reproduced in Fig.~\ref{pd_BS}.

\section{Conclusions}
\label{sec:Conclusions}

In this paper we analyzed the  Raman response of an AFe$_2$Se$_2$ superconductor assuming that the symmetry of the superconducting state is the ``other'' $s^{+-}$ 
 state \cite{Mazin2011,Khodas2012a}, in which  the gap is s-wave, but it changes sign between the two hybridized electron pockets.
 We focused on Raman response in $B_{2g}$ channel in the actual 2Fe BZ.

We found that $B_{2g}$ Raman susceptibility at $T=0$ exhibits the double-peak structure, Fig.~\ref{fig:Raman_1}.
The two peaks correspond to two distinct in-gap $B_{2g}$ symmetric collective modes.
The first mode is the BS mode in the Cooper channel, and its existence is due to the fact that the pairing interaction in the $d-$wave channel is 
  weaker than that in $s^{+-}$ channel, but nevertheless is attractive.
The second mode is the exciton in the particle-hole channel. This mode emerges because density-density interaction in $B_{2g}$ channel is also attractive. 
The $d-$wave attraction emerges from the original Hubbard-type repulsion because  density-density interaction in the $B_{2g}$ channel changes sign between the two hybridized electron pockets.
This sign reversal is akin to the transformation of the Hubbard repulsion into a attraction in $s^{+-}$  Cooper channel.
This situation should be contrasted with that in a single band superconductors where the interaction in the particle-hole  channel is in general a repulsive one.

In a generic situation the BS mode and particle-hole exciton are strongly mixed in which case only a single undamped in-gap mode survives, 
  the other is pushed  above $2\Delta$ threshold (see  Fig.~\ref{fig:Raman_BCS}).

This does not happen for $s^{+-}$ superconductor as the vertex which couples particle-particle and particle-hole channels is an odd function of an $s^{+-}$ gap, and the 
 contributions to this vertex bonding and anti-bonding Fermi pockets nearly cancel each other, the net result remains finite only due to a finite ellipticity of electron pockets.
As a result, both modes remain below $2 \Delta$ and the Raman intensity $\chi''(\omega)$ has two distinct peaks,  Fig.~\ref{fig:modes}.
 The decoupling between the two channels becomes exact at the boundary between $s^{+-}$  and $s+id$ phases, at  $\kappa = \kappa_2$ along $T=0$ line on the phase diagram in Fig.~\ref{pd_BS}.

\begin{figure}[h]
\begin{center}
\includegraphics[width=1.0\columnwidth]{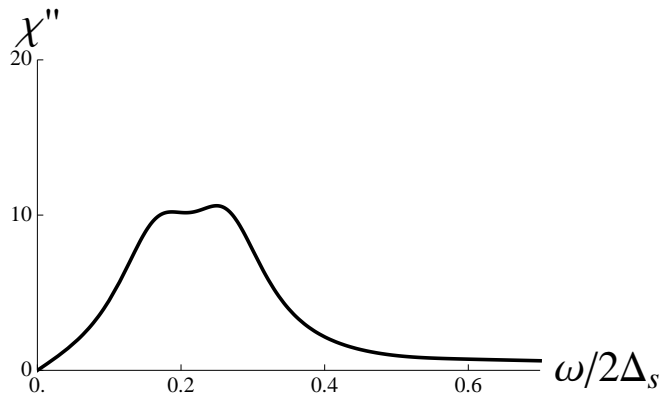}
\caption{
Calculated Raman susceptibility, $\chi''$ as a function of the dimensionless frequency, $\omega/ 2 \Delta_s$ for the fixed $\kappa=0.63 > \kappa_2$.
The two peaks represent in-gap modes corresponding to the two solid lines in Fig.~\ref{fig:modes}.
The parameters used in the calculation are $u_d = 0.4$,  $u_{\rho}=0.98$, $\mu_+/\mu_- = 0.05$.
The finite imaginary part $0.07 (2 \Delta_s) \ll 2 \Delta_s$ is added to the frequency to simulate the effect of the disorder induced smearing.}
\label{fig:exp}
\end{center}
\end{figure}

We compared our results with $B_{2g}$ Raman data for K$_{0.75}$Fe$_{1.75}$Se$_2$, reported in Ref.~\cite{Ignatov2012}.
The double-peak structure of $\chi''(\omega)$ combined with inhomogeneous broadening gives rise to Raman profile with intensity enhanced in a finite frequency window below $2\Delta$ (see Fig.~\ref{fig:exp}). We argue that this is quite consistent with the data.
We note that the interval between the two peaks can be filled with the $B_{2g}$ intensity due to the higher order processes originating form non-linear mode coupling, \cite{Barlas2013}.

\begin{acknowledgements}
We  thank R.M. Fernandes, A. Levchenko M.G. Vavilov for useful discussions, A. Ignatov for the help with experiments, and N.L.~Wang for K$_{0.75}$Fe$_{1.75}$Se$_{2}$ crystals.
M.K. acknowledges support by the University of Iowa.
The work by A.V.C. was supported by the Office of Basic Energy Sciences U.S. Department
of Energy, Division of Materials Sciences and Engineering, under the Award \#DE-FG02-ER46900.
The work by G.B. was supported by the US Department of Energy, Office of Basic Energy Sciences,
Division of Materials Sciences and Engineering, under the Award \#DE-SC0005463.
\end{acknowledgements}

\begin{appendix}

\section{Calculation of the polarization operators}
\label{app:Pi}
In this section we give details of calculation of the polarization operators as defined by the Eq.~\eqref{Pi}.
The calculation of diagonal polarization operators, $\Pi_{22}$ and $\Pi_{33}$ differs  from the calculation of the off-diagonal polarization operators, $\Pi_{23}$ and $\Pi_{32}$ in two respects.
First the diagonal polarization operators are require regularization at the ultra-violet, while the off diagonal polarization converge well enough to make it possible to integrate over the momentum and energy in arbitrary order.

The second difference is that while for the diagonal polarization operators the difference in the density of states is inessential, it is crucial in the case of the off diagonal polarization operators.
Hence we consider the two cases separately.

\subsection{Diagonal polarization operators, $\Pi_{22}$ and $\Pi_{33}$}
\label{app:Pi22}

In all the integrations here the small difference in the density of states for the subbands $a$ and $b$ is neglected and both species then contribute equally.
This results in extra factor of 2 compared for a single band case.
We start with the calculation of $\Pi_{22}$,
We decompose it into 
\begin{align}\label{app_pol_1}
\Pi_{22} (x) = \Pi_{22} (0) + [ \Pi_{22} (x) - \Pi_{22} (0) ]
\end{align}
where the first peace contains a logarithmic ultraviolet divergence, while the second piece is well convergent, and can be easily evaluated to give,
\begin{align}\label{app_pol_2}
\Pi_{22} (x) - \Pi_{22} (0) = -
4 x \left\langle  \frac{ c^2 \arcsin \left(x/s\right)}{\sqrt{s^{2}-x^2 }} \right\rangle\, .
\end{align}
The first, static term in \eqref{app_pol_1} reads
\begin{align}\label{pi_static}
\Pi_{22} (0) = -4 \left\langle c^2 \log\frac{2 \Lambda}{ s \Delta_s} \right\rangle\, ,
\end{align}
where $\Lambda$ is the ultraviolet cutoff.
To eliminate it in favor of coupling constants we consider the self consistency equation on the $s$-wave order parameter in the $s^{+-}$ phase, $\kappa> \kappa_2(T=0)$,
\begin{align}\label{self_s}
\frac{ 1 }{ u_s }  -  \int d \xi \int \frac{ d \epsilon}{ 2 \pi }
\left\langle \frac{ s^2 }{ \epsilon^2 + \xi^2 +s^2 \Delta_s^2 }  \right\rangle = 0\, .
\end{align}
Similarly to Eq.~\eqref{pi_static}, Eq.~\eqref{self_s} gives
\begin{align}\label{self_s1}
\frac{1}{ u_s } - 2 \langle s^2 \log \frac{ 2 \Lambda }{ \Delta_s s }\rangle=0
\end{align}
Equations \eqref{self_s1} yields
\begin{align}\label{self_s2}
\log \frac{ 2 \Lambda }{ \Delta_s }
=
\frac{\langle s^2 \log s^2 \rangle }{ 2 \langle s^2 \rangle} +
\frac{ 1 }{ 2 u_s \langle s^2 \rangle }
\end{align}
Substituting Eq.~\eqref{self_s2} to Eq.~\eqref{pi_static} we obtain
\begin{align}\label{pi_static1}
\Pi_{22}(0) = - \frac{2}{ u_s }\frac{ \langle c^2 \rangle}{ \langle s^2 \rangle}
-2 \frac{ \langle c^2\rangle}{ \langle s^2\rangle} \langle s^2 \log s^2 \rangle + 2\langle c^2 \log s^2 \rangle
\end{align}
Equations \eqref{app_pol_1}, \eqref{app_pol_2} and \eqref{pi_static1} yield Eq.~\eqref{pol22} of the main text.

We now turn to the calculation of the $d$-wave density polarization operator.
Similar to Eq.~\eqref{app_pol_1} we write
\begin{align}\label{pol33_1}
\Pi_{33} (x) = \Pi_{33} (0) + [ \Pi_{33} (x) - \Pi_{33} (0) ]\, .
\end{align}
The rationale for the decomposition, Eq.~\eqref{pol33_1} is that the second term is well convergent at the ultra-violet and the momentum integration can be performed first with the result,
\begin{align}\label{pol33_2}
\Pi_{33} (x) - \Pi_{33} (0) =
4 - 4 \left\langle
\frac{ c^2 s^2 \arcsin \left( x/s \right)}
{x \sqrt{s^2 - x^2}}
\right\rangle \, .
\end{align}
The static part gives the density of states,
\begin{align}\label{pol33_3}
\Pi_{33} (0) = - 4\, .
\end{align}
Equations \eqref{pol33_1}, \eqref{pol33_2} and \eqref{pol33_3} reproduce Eq.~\eqref{pol33} of the main text.

\subsection{Off diagonal polarization operators, $\Pi_{23}$ and $\Pi_{32}$}
\label{app:Pi23}
To evaluate $\Pi_{23}$ we substitute the representation, Eq.~\eqref{matrix_G2}
in the definition, Eq.~\eqref{Pi}.
We obtain after taking the trace over the subband indices,
\begin{align}\label{app_Pi_1}
&\Pi_{23}(\omega) = \int \frac{ d\epsilon}{ 2 \pi} \int \frac{ d^2 \vec{k}}{ (2 \pi)^2 }
c_{\vec{k}}^2 \mathrm{Tr}\left[ \tau_2
G_+(\epsilon + \omega)
\tau_3
G_+(\epsilon) \right]
\notag \\
&+   \int \frac{ d\epsilon}{ 2 \pi} \int \frac{ d^2 \vec{k}}{ (2 \pi)^2 }
c_{\vec{k}}^2 \mathrm{Tr}\left[ \tau_2
G_-(\epsilon + \omega)
\tau_3 G_-(\epsilon) \right] \, .
\end{align}
Taking the trace over Nambu indices in the first term of Eq.~\eqref{app_Pi_1} and using Eq.~\eqref{matrix_G2} we write,
\begin{align}\label{app_Pi_2}
& \int \frac{ d\epsilon}{ 2 \pi} \int \frac{ d^2 \vec{k}}{ (2 \pi)^2 }
\mathrm{Tr}\left[ \tau_2
G_+(\epsilon + \omega)
\tau_3
G_+(\epsilon) \right]
\notag \\
= &
-2 i \int \frac{ d\epsilon}{ 2 \pi} \int \frac{ d^2 \vec{k}}{ (2 \pi)^2 }
c_{\vec{k}}^2 \frac{  i (\epsilon + \omega)  + \xi^a_{\vec{k}} }{ (\epsilon + \omega)^2 + (\xi^a_{\vec{k}})^2 + s_{\vec{k}}^2 \Delta_s^2}
\notag \\
& \times
\frac{ s_{\vec{k}} \Delta_s  }{ \epsilon^2 + (\xi^a_{\vec{k}})^2 + s_{\vec{k}}^2 \Delta_s^2}
- \left( a \rightarrow b \right)\, .
\end{align}
The cross-product terms which contain both subband energies, $\xi^{a}_{\vec{k}}$ and $\xi^b_{\vec{k}}$ are omitted in the right hand side of the Eq.~\eqref{app_Pi_2}.
Such terms are not singular at $\omega \sim 2 \Delta_s$ and we shoud discard them in view of the approximation made in going from Eq.~\eqref{matrix_G} to Eq.~\eqref{matrix_G1}.
We note however that contributions of this kind from the two terms of Eq.~\eqref{app_Pi_1} cancel each other identically out.
The second term of Eq.~\eqref{app_Pi_1} is readily shown to make the contribution identical to that of Eq.~\eqref{app_Pi_2}.
Therefore we have
\begin{align}\label{app_Pi_3}
\Pi_{23} &=  - 4i \int \frac{ d\epsilon}{ 2 \pi} \int \frac{ d^2 \vec{k}}{ (2 \pi)^2 }
c_{\vec{k}}^2 \frac{  i (\epsilon + \omega)  + \xi^a_{\vec{k}} }{ (\epsilon + \omega)^2 + (\xi^a_{\vec{k}})^2 + s_{\vec{k}}^2 \Delta_s^2}
\notag \\
& \times
\frac{ s_{\vec{k}} \Delta_s  }{ \epsilon^2 + (\xi^a_{\vec{k}})^2 + s_{\vec{k}}^2 \Delta_s^2}
- \left( a \rightarrow b \right)\, .
\end{align}

Clearly the polarization operator in Eq.~\eqref{app_Pi_3} is non-zero only when $\xi^a \neq \xi^b$.
These energies are made unequal by a finite hybridization and/or ellipticity.
The expression for $\Pi_{23}$ is therefore non-universal, and depend on the fine details of the band structure.
We therefore do not attempt to consider it in full generality.
The only important message for us is that $\Pi_{23}$ is non-zero in generic situation.
For completeness and illustration we evaluate it for the model specified by Eqs.~\eqref{Ec} and \eqref{xi} such that
\begin{align}\label{app_Pi_4}
\xi^{a,b}_{\vec{k}} = \frac{k^2}{2 \mu_+} \pm \left( \lambda^2 + \frac{ k^4}{ 4 \mu_-^2 } \cos^2 2 \phi \right)^{1/2} - E_F\, .
\end{align}

We transform the momentum integration in Eq.~\eqref{app_Pi_3} following the standard prescription,
\begin{align}\label{change}
\int \frac{d^2 \vec{k}}{(2 \pi)^2 } f(\vec{k})
=
\int_0^{2 \pi} \frac{ d \phi }{ 2 \pi } N_{a,b} (\phi) f(\xi_{a,b},\phi)\, ,
\end{align}
where the densities of states
\begin{align}\label{N}
N_{a,b} = (2\pi)^{-1} \left(\frac{d \xi_{a,b}}{ k_{a,b} d k_{a,b} }\right)^{-1}
\end{align}
are slightly different.

The non-zero contribution to $\Pi_{23}$ arises from the two sources.
The first is due to the difference in the density of states, $\delta N = N_a - N_b$, and the second is due to the variation of the momentum dependent prefactors $c_{\vec{k}}^2 s_{\vec{k}}$ for the finite difference, $\delta \xi =\xi_a -\xi_b$.
Correspondingly we write
\begin{align}\label{app_Pi_3A}
\Pi_{23} = \delta_1 \Pi_{23} + \delta_2 \Pi_{23}\, .
\end{align}
In explicit form we have
\begin{align}\label{delta1}
\delta_1 \Pi_{23} &=
  - 4 i \int \frac{ d\epsilon}{ 2 \pi} \int d \xi \left\langle \delta N
c_{\vec{k}}^2 \frac{  i (\epsilon + \omega)  + \xi }{ (\epsilon + \omega)^2 + \xi^2 + s_{\vec{k}}^2 \Delta_s^2}
\right.
\notag \\
& \times
\left.
\frac{ s_{\vec{k}} \Delta_s  }{ \epsilon^2 + \xi^2 + s_{\vec{k}}^2 \Delta_s^2} \right\rangle\, ,
\end{align}
\begin{align}\label{delta2}
\delta_2 \Pi_{23} &=
  - 4i \int \frac{ d\epsilon}{ 2 \pi} \int d \xi \left\langle
  \partial_{\xi} (c_{\vec{k}}^2 s_{\vec{k}}) \delta \xi
 \frac{  i (\epsilon + \omega)  + \xi }{ (\epsilon + \omega)^2 + \xi^2 + s_{\vec{k}}^2 \Delta_s^2}
\right.
\notag \\
& \times
\left.
\frac{  \Delta_s  }{ \epsilon^2 + \xi^2 + s_{\vec{k}}^2 \Delta_s^2} \right\rangle\, .
\end{align}

We start with the contribution, $\delta_1 \Pi_{23}$, Eq.~\eqref{delta1}.
From Eq.~\eqref{app_Pi_4} we obtain
\begin{align}\label{app_Pi_5}
\frac{d \xi_{a,b}}{ k_{a,b} d k_{a,b} } \approx
\frac{1}{\mu_+} \pm \frac{k_F^2 \cos^2 2 \phi}{2 \mu_-^2} \left( \lambda^2 + \frac{ k^4_F }{ 4 \mu_-^2 }\cos^2 2 \phi  \right)^{-1/2}
\end{align}
with the definition, $k_F = \sqrt{ 2 m_+ E_F}$.
To the same accuracy, Eqs.~\eqref{N} and \eqref{app_Pi_5} yield
\begin{align}\label{app_Pi_6}
N_{a,b} \approx N_0\left( 1 \mp  \frac{ (\mu_+/ \mu_-) (k_F^2/ 2 \mu_-)\cos^2 2 \phi}{( \lambda^2 + (k_F^2/2 \mu_-)^2 \cos^2 2 \phi)^{1/2}}\right)\, ,
\end{align}
where the average density of states $N_0 = (2 \pi)^{-1} M$ is absorbed in our definitions of the scattering amplitudes, and is henceforth omitted.
In terms of the rotation angle in the orbital space, $\theta_{\vec{k}}$ defined by Eq.~\eqref{angle}
\begin{align}\label{app_Pi_7}
\delta N =  - 2\frac{ \mu_+}{ \mu_- } \frac{c_{\vec{k}}^2}{s_{\vec{k}}}\, ,
\end{align}
where we used the explicit expressions for the functions $c_{\vec{k}}$ and $s_{\vec{k}}$
\begin{align}\label{cs_explicit}
c_{\vec{k}} = \frac{   \cos 2 \phi }{ \sqrt{ \kappa^2 +  \cos^2 2 \phi} }
\, , \quad
s_{\vec{k}} = \frac{ \kappa }{ \sqrt{ \kappa^2 + \cos^2 2 \phi} }\, .
\end{align}
obtained by substitution of Eq.~\eqref{Ec} to the Eq.~\eqref{angle}.

To evaluate the contribution, $\delta_2 \Pi_{23}$, Eq.~\eqref{delta2} we note that
\begin{align}
\delta \xi  \partial_{\xi}\left[ (c_{\vec{k}})^2 s_{\vec{k}} \right]\approx 2 \mu_+ \delta \xi  \partial_{k^2} \left[(c_{\vec{k}})^2 s_{\vec{k}} \right]\, .
\end{align}
Using the explicit expressions \eqref{app_Pi_4} and \eqref{cs_explicit}
\begin{align}\label{delta_xi}
\delta \xi  \partial_{\xi} \left[(c_{\vec{k}})^2 s_{\vec{k}}\right]
=
- 2 \frac{\mu_+}{\mu_-}
\frac{ \kappa  \left(\cos^4\phi-2 \kappa ^2 \cos^2\phi \right)}
   {\left(\kappa ^2+\cos^2\phi\right)^2}\, .
\end{align}
We can write Eq.~\eqref{delta_xi} using Eq.~\eqref{cs_explicit} as
\begin{align}\label{delta_xi1}
\delta \xi  \partial_{\xi}\left[ (c_{\vec{k}})^2 s_{\vec{k}}\right]
=
- 2 \frac{\mu_+}{\mu_-}\kappa \left( c^4_{\vec{k}} - 2 s_{\vec{k}}^2 c_{\vec{k}}^2 \right)\, .
\end{align}

Combining Eqs.~\eqref{app_Pi_3A}, \eqref{delta1}, \eqref{delta2}, \eqref{app_Pi_7} and \eqref{delta_xi1} we write \eqref{app_Pi_3} in the form,

Equation \eqref{app_Pi_7} allows us to write Eq.~\eqref{app_Pi_3} in the form,
\begin{align}\label{app_Pi_8}
\Pi_{23}& =
-8 i \int \frac{ d\epsilon}{ 2 \pi} \int d \xi \left\langle {\cal F}(\mu_+/\mu_-;\kappa,\phi)
\right.
 \notag \\
  \times &
\frac{  i (\epsilon + \omega)  + \xi}{ (\epsilon + \omega)^2 + \xi^2 + s_{\vec{k}}^2 \Delta_s^2}
\left.
\frac{  \Delta_s  }{ \epsilon^2 + \xi^2 + s_{\vec{k}}^2 \Delta_s^2}\right\rangle\, .
\end{align}
Integration of Eq.~\eqref{app_Pi_8} over  $\xi$ and $\epsilon$ variables yields Eq.~\eqref{pol23}.
In the dispersion model specified by Eqs.~\eqref{H2} and \eqref{Ec} we
\begin{align}\label{F_cal}
{\cal F}\left(\frac{\mu_+}{\mu_-} \ll 1;\kappa,\phi \right) \approx   \frac{\mu_+}{\mu_-}
\left( 2c_{\vec{k}}^2 s_{\vec{k}}^2 - c_{\vec{k}}^4(1 + \kappa)    \right)\, ,
\end{align}
which is the equation \eqref{F_calig1} of the main text.

\section{Interaction amplitudes in the p-h channel}
\label{app:Interaction}
In this section we focus on the p-h channel of the generic interaction, Eq.~\eqref{Hint}.
The p-h channel in turn is decomposed into the spin-single and spin-triplet component.
In the absence of spin-orbit interaction we expect that these two channels are not mixed, and Raman probes the spin-singlet, i.e. density excitations.

We consider the decomposition of the first term $H_1$ in Eq.~\eqref{Hint} in details and henceforth quote the results for other three parts of the interaction Hamiltonian.
We start with singling out the direct and exchange terms of the interaction,
\begin{align}\label{H1_split}
H_1 = H_1^{dir} + H_1^{ex}\, ,
\end{align}
where the Cooper channel is omitted and
\begin{align}
H_1^{dir} &= u_1 \sum_{k k'} c_{\sigma k}^{\dag} c_{\sigma k}  f_{\sigma' k'}^{\dag} f_{\sigma' k'}
\notag \\
H_1^{ex} &= u_1 \sum_{k k'} c_{\sigma k}^{\dag} c_{\sigma k'}  f_{\sigma' k'}^{\dag} f_{\sigma' k}
\end{align}
To facilitate the decomposition of both parts of the Hamiltonian into density and spin channels we introduce intra-band density operators,
\begin{align}\label{rhoA}
\rho_c &= \sum_{k} c_k^{\dag} c_k \, ,\,\,\, \rho_f = \sum_{k} f_k^{\dag} f_k \,
\end{align}
and similarly, two inter-band density operators,
\begin{align}\label{rhoB}
\rho_{cf} &= \sum_{k} c_k^{\dag} f_k\, ,\,\,\,
\rho_{fc} = \sum_{k} f_k^{\dag} c_k \, .
\end{align}
It is then expedient to rearrange the density operators, Eq.~\eqref{rhoA} and \eqref{rhoB} into a symmetric and anti-symmetric combinations,
\begin{align}\label{rhoC}
\rho_{\pm} = \frac{1}{2} \left( \rho_c \pm \rho_f \right)\, , \quad
\rho_{\pm}^{\vec{Q}} =  \frac{1}{2} \left( \rho_{cf} \pm \rho_{fc}\right)\, ,
\end{align}
The symmetric (anti-symmetric) combinations are selectively excited by a photon in A$_{1g}$ (B$_{1g}$) Raman configurations respectively.
Hence, the total density $\rho_+$ is not involved in the case of $B_{1g}$ Raman  configuration we study in this paper.
We will see below that the only relevant anti-symmetric density combination is $\rho_-$.
This is clearly a nematic density operator proportional to the difference between the population of the two pockets in 1 Fe BZ.

In terms of the densities introduced above,  Eq.~\eqref{rhoA} we have
\begin{align}\label{H1_dir}
H_1^{dir} = u_1 \rho_c \rho_f \, .
\end{align}
To rewrite the exchange Hamiltonian in a similar fashion, we introduce the intra- and inter-band spin density components parallel to the definitions \eqref{rhoA},\eqref{rhoB} and \eqref{rhoC} via,
\begin{align}
\vec{S}_{c} & = \frac{1}{2}\sum_k c_{\sigma k }^{\dag} \vec{\sigma}_{\sigma \sigma'} c_{\sigma' k }\, , \,\,\,
\vec{S}_{f}  = \frac{1}{2}\sum_k f_{\sigma k }^{\dag} \vec{\sigma}_{\sigma \sigma'} f_{\sigma' k }
\end{align}
and similarly,
\begin{align}
\vec{S}_{cf} = \frac{1}{2}\sum_k c_{\sigma k }^{\dag} \vec{\sigma}_{\sigma \sigma'} f_{\sigma' k }\, ,
\end{align}
with $\vec{S}_{fc}$ obtained from $\vec{S}_{cf}$ by the interchange $c \leftrightarrow f$.
With the above definitions the exchange part takes the form,
\begin{align}\label{H1_ex}
H_1^{ex} =
- u_1 \left[ 2 \vec{S}_{cf} \cdot \vec{S}_{fc} + \frac{1}{2} \rho_{cf} \rho_{fc} \right]\, .
\end{align}

Combining Eqs.~\eqref{H1_split}, \eqref{H1_dir} and \eqref{H1_ex} we have for the first term of Eq.~\eqref{Hint}
\begin{align}
H_1 = u_1 \left[ \rho_c \rho_f  - 2  \vec{S}_{cf}\cdot \vec{S}_{fc} - \frac{1}{2} \rho_{cf}\rho_{fc} \right]\, .
\end{align}
Following essentially the same steps we obtain
\begin{align}
H_2 = u_2 \left[ \rho_{cf} \rho_{fc} - 2 \vec{S}_c \cdot \vec{S}_f - \frac{1}{2}   \rho_c \rho_f \right]\, ,
\end{align}
\begin{align}
H_3 =  &
\frac{u_3}{4} \left( \rho_{cf} \rho_{cf}  +  \rho_{fc} \rho_{fc} \right)
\notag \\
& -
u_3 \left( \vec{S}_{cf} \cdot \vec{S}_{cf} + \vec{S}_{fc} \cdot \vec{S}_{fc} \right)\, ,
\end{align}
and
\begin{align}
H_4 = u_4 \left[
\frac{1}{4} \left( \rho_{c} \rho_{c} \! + \! \rho_{f} \rho_{f} \right)  -
\left( \vec{S}_{c} \cdot \vec{S}_{c}\! +\! \vec{S}_{f} \cdot \vec{S}_{f} \right) \right].
\end{align}
Here we are interested in the density (singlet) part of the total Hamiltonian which reads
\begin{align}\label{H_ch1}
H_{ch} &= \frac{u_4}{ 4 } \left[ \rho_c \rho_c + \rho_f \rho_f \right]
+
\left( u_1 - \frac{ u_2}{ 2} \right) \rho_c \rho_f
\notag \\
+&
\left( u_2 - \frac{ u_1}{ 2} \right) \rho_{cf} \rho_{fc}
+
\frac{u_4}{2}\left( \rho_{cf} \rho_{cf} + \rho_{fc} \rho_{fc} \right).
\end{align}
We now rewrite Eq.~\eqref{H_ch1} in terms of the symmetric combinations \eqref{rhoC} as follows
\begin{align}\label{H_ch2}
H_{ch} &= \left[ \frac{u_4}{2} + \frac{u_2}{2} - u_1 \right] \rho_- \rho_-
 \\
+ &
\left[ \frac{u_4}{2} - \frac{u_2}{2} + u_1 \right] \rho_+ \rho_+
\notag \\
+ &
\left[ \frac{u_3}{2}  - \frac{u_1}{2} + u_2  \right] \rho_+^{\vec{Q}} \rho_+^{\vec{Q}}
+
\left[\frac{u_3}{2} + \frac{u_1}{2} - u_2\right] \rho_-^{\vec{Q}} \rho_-^{\vec{Q}}\, \notag .
\end{align}
The second term of Eq.~\eqref{H_ch2} is not important in B$_{1g}$ configuration.
The remaining $d$-wave combination $\rho_-^Q \rho_-^Q$ lacks intra-band components in the hybridized $ab$ basis and is neglected under the conditions stated in Sec.~\ref{sec:Model}.
Hence, the Eq.~\eqref{V_rho_d} of the main text is obtained with the amplitude $u_{\rho}$ given by Eq.~\eqref{amplitudes}.
To explore the possibility of attraction in the $d$-wave density channel we rewrite the corresponding interaction amplitude in terms of the Hubbard and Hund interaction amplitudes using \cite{Khodas2012a}
\begin{align}
u_1 = U - J\, , u_2 = - 2 J - J'\, , u_4 = U - 3 J\, .
\end{align}
Therefore the charge channel interaction, Eq.~\eqref{H_ch2}
\begin{align}
H_{ch} =  -\frac{1}{2} \left[ U + 3 J + J' \right] \rho_- \rho_-\, ,
\end{align}
and may become attractive ($u_{rho}<0$) if the Hund couplings $J$ and $J'$ are large.

\section{Collective modes in $d$-wave symmetry channel in BCS theory}
\label{app:BCS}

In this appendix we review the BS-type excitations in the coupled particle-hole and Cooper channels for a single band material.
The BS modes were first studied in \cite{Bardasis1961}.
The results of that work can be summarized as follows.
For each symmetry channel defined by a total angular momentum $L$ and its projection on a given axes $M$ the two types of collective modes were identified.
 These excitations can be envisioned as the oscillations of the real  and imaginary  parts of the complex $L \neq 0$ OP, taking the dominant $L=0$, $s$-wave OP to be real.
The above two eigen-mode oscillations are referred to as $\Lambda_{LM}$ (``amplitude'') and $\Gamma_{LM}$ (``phase'') modes respectively.
The in-phase, $\Lambda_{LM}$, ``amplitude'' modes are the oscillations of the real part of the $d$-wave OP assuming $\Delta_s$ is real.
The oscillations of the imaginary part of the $d$-wave OP,  gives rise to the $\Gamma_{LM}$ mode.
The distinction between the two modes of oscillations is due to a non-zero $s$-wave OP in the ground state.
It is only the overall $U(1)$ gauge freedom which is preserved, while the action explicitly depends on the relative phase of $s$ and $d$-wave OPs.
 BS have shown  that the $\Gamma_{LM}$ is bound below $2 \Delta$ and is undamped provided the interaction in the $LM$ channel is attractive, while the $\Lambda_{LM}$ falls into the quasi-particle continuum and is damped.
In a superconductor the Cooper and p-h channels are fundamentally interrelated, and in general should be considered on equal footing.
The effect of the coupling between these two channels is most dramatic in the dominant, $L=0$ channel -- the long range Coulomb interaction lifts the the otherwise gapless phase-oscillations up to a plasma frequency.
In the subdominant channels with $L\neq 0$ the effect of long-range Coulomb interaction is much less pronounced and the interaction can often be approximated
 as local.
In fact, BS have demonstrated  that long-range Coulomb repulsion has a negligible effect on the binding energy of $L\neq 0$ excitons.

Below we assume that the ground state is $s$-wave symmetric with simple isotropic gap, $\Delta_s$, and consider the excitations in the $d$-wave channel.

The observables describing $\Gamma$ ($\Lambda$) and  fluctuations are represented by Nambu bilinears as
\begin{align}\label{struc1}
\chi^{\dag}_{\vec{k}}  \tau_{1} \chi_{\vec{k}}  &=  Y_{LM}(\hat{\vec{k}})
\left(
\psi_{-\vec{k}\downarrow} \psi_{\vec{k}\uparrow} +
 \psi^{\dag}_{\vec{k}\uparrow} \psi^{\dag}_{-\vec{k}\downarrow}\right)  \notag \\
\chi^{\dag}_{\vec{k}}  \tau_{2} \chi_{\vec{k}}  & =
 Y_{LM}(\hat{\vec{k}}) i
\left(
\psi_{-\vec{k}\downarrow} \psi_{\vec{k}\uparrow} -
 \psi^{\dag}_{\vec{k}\uparrow} \psi^{\dag}_{-\vec{k}\downarrow}\right)  \, ,
\end{align}
where the Nambu spinors are $\chi^{\dag}_{\vec{k}} = \left[ \psi_{\vec{k}\uparrow}^{\dag},\psi_{-\vec{k}\downarrow} \right]$, and spherical harmonics $Y_{LM}$ encode the angular dependence of the Cooper pair exciton wave-function, \cite{Bardasis1961}.

The BS modes dispersion, $\omega_{\Lambda,\Gamma}(\vec{q})$ are found from the equations
\begin{align}\label{LG}
u_d^{-1} + \Pi_{11,22}(\omega_{\Lambda,\Gamma}(\vec{q}),\vec{q}) = 0\, .
\end{align}
The polarization operators entering \eqref{LG} are straightforwardly calculated at $\vec{q}=0$, \cite{Bardasis1961},
\begin{subequations}
\begin{align}\label{PI_11}
\Pi_{11}(x,0) =  - u_s^{-1}
+
 2 \left(\frac{\arcsin\left( x \right)}{x}\right) \sqrt{1-x^2 } \, ,
\end{align}
\begin{align}\label{PI_22}
\Pi_{22}(x,0) = - u_s^{-1}
- 2 x \frac{  \arcsin\left( x \right)}{\sqrt{1-x^2 }} \, ,
\end{align}
\end{subequations}
where $x = \omega / 2 \Delta_s$.
The equation \eqref{LG} has no real solutions  for $\omega_{\Lambda}$, and the ``amplitude'' $\Lambda$ mode falling in the quasi-particle continuum is damped, \cite{Bardasis1961}.
In contrast, the ``phase'' $\Gamma$ mode is long-lived in-gap excitation.
The divergence of $\Pi_{22}$ at the kinematical threshold, $x = 1$ guarantees that the solution $\omega_{\Gamma}<2 \Delta_s$ of Eq.~\eqref{LG}, the BS mode exists.

The charge exciton is obtained if the interaction in the charge sector is attractive.
The energy, $\omega_{ex}$ of the exciton is determined by the equation similar to \eqref{LG}
\begin{align}\label{exciton}
u_{\rho}^{-1} + \Pi_{33}(\omega_{ex}(\vec{q}),\vec{q}) = 0\, ,
\end{align}
where $u_{\rho}$ is the scattering amplitude in the density channel.
The polarization operator $\Pi_{33}(\omega,0)$, defined by \eqref{Pi} reads
\begin{align}\label{PI_33}
\Pi_{33}(x,0)
=
-2 \frac{\arcsin(x)}
   {x \sqrt{1- x^2}}
\end{align}
and has the divergence at $x =1$ similar to \eqref{PI_22}, and Eq.~\eqref{exciton} has a bound state exciton solution.

The coupling of the charge and Cooper channels is expressed as an of diagonal polarization operator.
In the BCS theory this polarization operator
\be\label{PI_23}
\Pi_{32}(x)= \Pi_{23}(x) = \frac{2  \arcsin\left( x\right)}
{\sqrt{1-x^2}}
\ee
is characterized by the same threshold divergence at $x = 1$ as Eq.~\eqref{PI_22} and \eqref{PI_33}.
For that reason the polarization operator, \eqref{PI_23} which couples the particle-hole and Cooper channels should be considered on equal footing with \eqref{PI_22} and \eqref{PI_33}.
On the other hand $\Pi_{31} = \Pi_{13}=0$ and the ``amplitude'', we the $\Lambda$ mode can be excluded from the consideration.

It follows that Eq.~\eqref{pol_op} reduces to Eqs.~\eqref{PI_22}, \eqref{PI_33} and \eqref{PI_23} in the limit of isotropic gap, $s_{\vec{k}} = c_{\vec{k}}$ up to an overall factor of $2$ due to the two pockets.

\begin{figure}[h]
\begin{center}
\includegraphics[width=1.0\columnwidth]{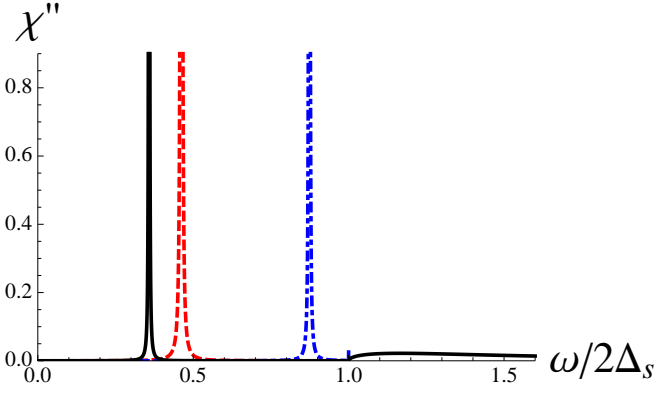}
\caption{ (color online)
Raman intensity, $\chi_{r,r}''$ as a function of $x = \omega / 2 \Delta_s$.
Solid (black) line is obtained by substitution of Eqs.~\eqref{PI_22}, \eqref{PI_33} and \eqref{PI_33} in Eq.~\eqref{I_R} for $u_s = 0.5$, $u_d =0.4$, and $u_{\rho} = 0.2$.
Dashed (red) line is calculated Raman intensity with only Cooper channel included, $u_{\rho}=0$.
Dashed (blue) line is calculated Raman intensity with only particle-hole channel included, $u_{d}=0$.
A small imaginary part is added to the frequency for illustration purposes, $x \rightarrow x + i \Gamma$, $\Gamma = 10^{-5}$.
The quasi-particle continuum is shown only for the full Raman susceptibility.
 }
\label{fig:Raman_BCS}
\end{center}
\end{figure}

To study the collective mode of coupled particle-hole and Cooper channels we define the $T$-matrix as
\begin{align}\label{T}
\hat{T}(x)=
\left( \hat{V}^{-1} + \hat{\Pi} (x)\right)^{-1}\, ,
\end{align}
where the polarization operator matrix
\begin{align}\label{Pi_2by2}
\hat{\Pi} =
\begin{bmatrix}
\Pi_{22} & \Pi_{23} \\
\Pi_{32} & \Pi_{33}
\end{bmatrix}\,
\end{align}
with the entries defined by Eqs.~\eqref{PI_22}, \eqref{PI_33} and \eqref{PI_23}.
The interaction matrix is diagonal, since it conserves the total number of particles,
\begin{align}
\hat{V} =
\begin{bmatrix}
u_d & 0 \\
0 & u_{\rho}
\end{bmatrix}\, .
\end{align}
Collective modes we are after are the solutions of
\begin{align}\label{det}
\det{\hat{T}(x)} = 0 \, .
\end{align}
We rewrite Eq.~\eqref{T} using the relations \eqref{PI_22}, \eqref{PI_33} and \eqref{PI_23} we get,
\begin{align}\label{T_1}
\hat{T}^{-1}(x) =
\begin{bmatrix} \delta u^{-1} & 0 \\ 0 & u_{\rho}^{-1}  \end{bmatrix}
- 2 \frac{ \arcsin x }{ \sqrt{ 1 - x^2 }  }\begin{bmatrix} x & -1 \\ -1 & x^{-1}
\end{bmatrix} \, ,
\end{align}
where
$ \delta u^{-1} = u_d^{-1} - u_{s}^{-1}$ is positive.
The matrix \eqref{T_1} can be explicitly diagonalized in two steps.
Consider the transformed matrix
\begin{align}\label{T_2}
\hat{T}'(x) = \hat{\Lambda}^{tr}  \hat{T}(x)  \hat{\Lambda}
\end{align}
with the choice
\begin{align}
\Lambda = \begin{bmatrix} \sqrt{ (\delta u^{-1})^{-1}}& 0 \\ 0 & \sqrt{u_{\rho}}  \end{bmatrix}
\end{align}
equation \eqref{T_2} becomes
\begin{align}\label{T_3}
\hat{T}'(x) = \hat{I}_2 + \hat{A}(x)
\end{align}
and it remains to diagonalize the matrix,
\begin{align}
\hat{A}(x) =
- 2 \frac{ \arcsin x }{ \sqrt{ 1 - x^2 }  }
\begin{bmatrix} x \delta u  & -\sqrt{\delta u u_{\rho}} \\ -\sqrt{\delta u u_{\rho}} & x^{-1} u_{\rho} \end{bmatrix}\, .
\end{align}
The resulting two eigenvalues of the transformed matrix \eqref{T_1} are
\begin{align}\label{lambda_1}
\lambda_1(x)  = 1 - 2 \frac{ \arcsin x }{ \sqrt{ 1 - x^2 }  } \left( x (\delta u^{-1})^{-1}  + x^{-1} u_{\rho}  \right)
\end{align}
and $\lambda_2 = 1$.
Due to the strong kinematical coupling between the two channels the would be two modes in Cooper and particle-hole channels merge into a single mode with the frequency $x_{res}$ satisfying $\lambda_1(x_{res}) = 0$.
In the weak coupling limit, $\{u_{\rho}, (\delta u^{-1})^{-1} \} \ll 1$ the amplitudes in two channels contribute additively to the binding energy,
\be
x_{res} \approx 1 -\frac{ \pi^2}{ 2}\left( (\delta u^{-1})^{-1} + u_{\rho} \right)^2 \, .
\ee
and in general are equally important.
The eigenvector of the $T$-matrix, \eqref{T} corresponding to the resonance frequency $x_{res}$ in this limit becomes,
$[- (\delta u^{-1})^{-1}, u_{\rho}]^t$.
It follows that the charge and Cooper channels participate in the coupled mode in proportion to their contribution to the binding energy a expected.

The exciton binding energy approaches $2 \Delta_s$ when the pairing amplitude in the $s$-wave channel exceeds the amplitude in the $d$-wave channel only slightly, $u_d \lesssim u_s$.
At the degeneracy limit $u_d = u_s$, the collective mode softens to zero,
$x_{res} = 0$, regardless of the interaction in the particle-hole channel, $u_{\rho}$.
Indeed, the two channels uncouple in the limit $x=0$ because at low frequencies $\Pi_{23}=\Pi_{32} =0 $ in this limit.
The physical reason for this is that only derivatives of the phase can enter the gauge invariant action.

We now consider the asymptotic softening of the collective mode in the degeneracy limit, $u_d \rightarrow u_s$, namely in the limit of small $\delta u^{-1}$.
Looking for the solution of $\lambda_1(x) = 0$ see Eq.~\eqref{lambda_1}, in the scaling form $x = y [\delta u^{-1} ]^{\alpha}$ with $y$ approaching a constant $y_0$ in the same limit we realize that $\alpha = 1/2$ and $y_0 = \sqrt{1 - u_{\rho}/2}$.
In other words,
\be\label{x_soft}
x_{res} \approx \sqrt{ \delta u^{-1} \left(1 - u_{\rho}/2 \right) }
\ee
asymptotically in the degeneracy limit, $u_d \rightarrow u_s$, $\delta u^{-1} \rightarrow 0$.
The eigenvalue of the $T$-matrix with the eigenvalue \eqref{x_soft} is
$\left[- \sqrt{ ( 1 - u_{\rho}/2)(\delta u^{-1})^{-1} }, \sqrt{u_{\rho}} \right]$ which means that this mode is predominantly made of $d$-wave Cooper pairs as again expected.

Equation \eqref{im_chi1} determines the Raman susceptibility, and we illustrate it for the case of isotropic gap in Fig.~\ref{fig:Raman_BCS}.

\end{appendix}


\end{document}